\documentclass[12pt,a4paper]{article}
\pdfoutput=1
\usepackage{jheppub}
\usepackage[sort&compress]{natbib}
\usepackage{url}
\usepackage{hyperref}
\usepackage{amsfonts}
\usepackage{epsfig}
\usepackage{cancel}
\usepackage{dcolumn}

\usepackage{amsmath}
\usepackage{amssymb}
\usepackage{amsthm}
\usepackage{amsfonts}
\usepackage{verbatim}
\usepackage{enumerate}
\usepackage{graphicx}
\usepackage{accents}
\usepackage{bm}

\usepackage{physics}
\usepackage{tensor}
\usepackage{siunitx}
\usepackage{circuitikz}
\usepackage{wasysym}

\usepackage{graphics}
\usepackage{caption}
\usepackage{subcaption}
\usepackage{epstopdf}
\usepackage[utf8]{inputenc} 
\usepackage{pgfplots}
\usepackage{slashed}

\newcommand{\SM}{\text{\tiny{SM}}}
\newcommand{\xiinf}{\xi_{\infty}}
\newcommand{\eq}{\text{eq}}

\newcommand{\etap}{{\bar{\eta}^{\prime}}}
\newcommand{\deltaN}{{\Delta}}
\newcommand{\deltaNb}{{\,\overline{\!\Delta\!}}}
\newcommand{\rs}{r_s}

\newcommand{\ccdot}{\!\cdot\!}
\renewcommand{\dv}[2]{\frac{ d #1}{d #2}}

\makeatletter
\g@addto@macro\bfseries{\boldmath}
\makeatother

\begin{document} 

\title{Large $N$-ightmare Dark Matter}

\author{Logan Morrison, Stefano Profumo, and}
\affiliation{Department of Physics and Santa Cruz Institute for Particle Physics\\University of California, Santa Cruz, CA 95064, USA}


\author{Dean J. Robinson}
\affiliation{Ernest Orlando Lawrence Berkeley National Laboratory, University of California, Berkeley, CA 94720, USA}

\emailAdd{loanmorr@ucsc.edu}
\emailAdd{profumo@ucsc.edu}
\emailAdd{drobinson@lbl.gov}

\abstract{A dark QCD sector is a relatively minimal extension of the Standard Model (SM) that admits Dark Matter (DM) candidates, but requires no portal to the visible sector beyond gravitational interactions: 
A ``nightmare scenario'' for DM detection.
We consider a secluded dark sector containing a single flavor of light, vector-like dark quark gauged under $SU(N)$. 
In the large-$N$ limit, this single-flavor theory becomes highly predictive, generating two DM candidates whose masses and dynamics are described by few parameters: 
A light quark-antiquark bound state, the dark analog of the $\eta'$ meson, and a heavy bound state of $N$ quarks, the dark analog of the $\Delta^{++}$ baryon. 
We show that the latter may freeze-in with an abundance independent of the confinement scale, 
forming DM-like relics for $N \lesssim 10$, while the former may generate DM via cannibalization and freeze-out.
We study the interplay of this two-component DM system,
and determine the characteristic ranges of the confinement scale, 
dark-visible sector temperature ratio, and $N$ that admit non-excluded DM,
once effects of self-interaction constraints and bounds on effective degrees of freedom at the BBN and CMB epochs are included.
}

\maketitle

\section{Introduction}

A comprehensive experimental program to search for the fundamental particle nature of the cosmological dark matter (DM) has been underway for decades~\cite{Bertone:2010zza}. 
Thus far there is no evidence for any non-gravitational DM signal. 
From the standpoint of cosmology and structure formation, the dark matter might well belong to a fully `secluded' dark sector, 
i.e., with no microscopic interactions with the Standard Model (SM) sector besides gravity: 
The so-called `nightmare scenario' for DM detection.
Absent non-gravitational interactions, probing such a dark sector is relegated to indirect consequences such as the shape of halos, or the impact of dark sector particles on the expansion rate of the universe.
Numerous models of secluded DM have been discussed in the literature (see for instance Ref.~\cite{Yang:2019bvg}); 
in light of ever-tightening DM detection bounds, scenarios in which dark matter interacts solely via gravitational interactions should be earnestly contemplated. 

A secluded dark sector gauged under a confining Yang-Mills theory may generate DM candidates in the form of various bound states,
without requiring a portal to the SM sector beyond gravitational interactions.
Strongly-coupled composite dark matter models have a relatively long history (for early work on this see Ref.~\cite{Carlson:1992fn}; for a recent comprehensive review, see Ref.~\cite{Kribs:2016cew}). 
Broadly, the phenomenology depends foremost on whether or not there exists a dark--SM portal, and secondly on the relative hierarchy between the confinement scale and the dark quark masses. 
Recent specific examples are presented in 
Ref.~\cite{Antipin:2015xia,Strassler:2006im,Kribs:2009fy,Antipin:2014qva,Appelquist:2014jch, Huo:2015nwa,Cline:2016nab,Berryman:2017twh, Hochberg:2014kqa, Hochberg:2015vrg}, 
that feature various degrees of complexity.

In this paper we study a relatively minimal secluded sector of this type, containing a single flavor of light vector-like dark quark gauged under a confining dark $SU(N)$: A single-flavor dark QCD.
Taking the large-$N$ limit~\cite{HOOFT1974461} (see e.g. Refs.~\cite{Coleman:1985rnk,Manohar:1998xv} for a review), the dynamics and spectrum of the confined theory become highly predictive, 
allowing one to develop a comprehensive picture of the cosmology and phenomenology of this type of dark sector
(the large-$N$ limit of a theory with \emph{heavy} dark quarks is discussed in Ref.~\cite{Mitridate:2017oky}).
The spectrum of the confined theory contains two stable states: A light pseudoscalar meson -- denoted `$\etap$', the dark analog of the $\eta'$ 
-- whose mass vanishes as $1/\sqrt{N}$ in the large-$N$ limit~\cite{Veneziano:1976wm,WITTEN1979269}; 
as well as a heavy baryon -- denoted `$\deltaN$', the dark analog of the $\Delta^{++}$ -- whose mass scales with $N$. 
The phenomenology of this secluded sector is mainly parametrized by the chiral symmetry breaking scale $\Lambda$, the rank $N$, 
the first two parameters of the momentum expansion of the chiral Lagrangian, and the temperature ratio of the dark--SM sector at early times. 
In particular, expanding in $1/N$, the leading-order terms of meson or baryon correlation functions are fully characterized by well-known topological arguments, 
allowing one to directly express relevant masses and interaction cross-sections solely in terms of these parameters, up to $\mathcal{O}(1)$ nuisance parameters. 

On the one hand, the heavy $\deltaN$ baryons -- rough analogs of skyrmions -- 
are pair-produced from the confined plasma only via exponentially suppressed-in-$N$ interactions, 
as first discussed in Ref.~\cite{Witten:1979kh}.
We observe that this creates a freeze-in DM candidate, whose relic abundance is exponentially sensitive only to $N$ but independent of $\Lambda$. 
If additional physics is present that permits $\etap$ to decay, e.g., to a dark photon, one can then generate extremely heavy $\deltaN$ DM.
On the other hand, the light $\etap$ undergo ``cannibalization'' controlled by their $4 \to 2$ annihilation, followed by a freeze-out.
(Because we consider only a single flavor, the 5-point Wess-Zumino-Witten term is absent -- for $SU(N)$ and $SO(N)$, 
this term requires at least three flavors, for $SP(N)$ at least two \cite{Kribs:2016cew} -- so that there are no $3 \to 2$ interactions.)
For earlier studies of cannibal dark matter scenarios see Refs.~\cite{Carlson:1992fn, deLaix:1995vi, Bernal:2015ova,Bernal:2015xba, Kuflik:2015isi, Soni:2016gzf, Hochberg:2014dra, Hochberg:2014kqa, Pappadopulo:2016pkp, Buen-Abad:2018mas, Erickcek:2020wzd}. 

We study the interplay of this two-component DM system in the large-$N$ limit, 
showing that the $\deltaN$ relic abundance can match the DM one for the range $N \lesssim 10$, 
while the $\etap$ produce DM according to a power-law $\Lambda \sim N^{-3/2}$.
Taken together as two-component DM system, the $\etap$ and $\deltaN$ components each produce a bound on the allowed range of confinement scale, $\Lambda$, and/or rank $N$ 
for such a system to produce the correct DM abundance.
We study this two-component system for a series of numerical benchmarks, characterizing the various allowed regimes.
Effects of relevant self-interaction constraints and bounds on effective degrees of freedom, $\delta N_{\text{eff}}$, at the BBN and CMB epochs 
imply that the early dark-SM temperature ratio must be quite small $\sim 10^{-2}$, which would suggest non-trivial dynamics in the early universe.
We show these bounds further imply that the most minimal scenario for this type of dark sector is ruled out, while near-minimal scenarios may be probed in the near future via DM self-interactions.

This paper is structured as follows. In Sec.~\ref{sec:LNQCD} we first review the details and leading order large-$N$ scaling of the dark sector spectrum and interactions. 
We then present in Sec.~\ref{sec:darkcos} an approximate analysis of the cosmological evolution of this two-component SM secluded sector, 
as well as the effects of relevant self-interaction and $\delta N_{\text{eff}}$ constraints. 
We proceed in Sec.~\ref{sec:numstudy} to implement a full numerical study, verifying and/or lifting various approximations from Sec.~\ref{sec:darkcos}, 
for several benchmark scenarios.
Section~\ref{sec:summ} summarizes our results.

\section{Large-$N$ single-flavor dark QCD}
\label{sec:LNQCD}
\subsection{$1/N$ expansion}
Our dark sector comprises a single flavor of vector-like `dark quarks', $q$ and $\bar{q}$, furnishing the (anti)fundamental of a dark $SU(N)$ gauge interaction
i.e. $q \sim \bm{N}$ and $\bar{q} \sim \overline{\!\bm{N}}$. The dark sector perturbative Lagrangian
\begin{equation}
    \mathcal{L}_{\mathrm{dark}} \supset -\frac{1}{4}G^{\mu\nu,a}G^{a}_{\mu\nu} +  i\bar{q}(\slashed{D} - m_q)q\,.
\end{equation}

As usual in large-$N$ constructions~\cite{HOOFT1974461,Coleman:1985rnk}, 
the $SU(N)$ gauge coupling is rescaled as $g/\sqrt{N}$ with 't\,Hooft coupling $g$ fixed, i.e. $D_\mu = \partial_\mu -i(g/\sqrt{N}) T^a A^a_\mu$.
This generates a well-controlled expansion in $1/N$, for $N \gg 1$. 
In particular, the coefficient of the gauge coupling, $\beta$-function $b_0 = 11/3 - 4/(3N)$, such that
the theory is expected to undergo confinement with a chiral symmetry breaking scale, $\Lambda$, that is independent of $N$ at leading order. 
The scale $\Lambda$ may thus be treated as an independent parameter of the theory in the large-$N$ limit.
Further, in the large-$N$ limit the dark quark mass remains a technically natural parameter: 
We consider the regime $m_q \lll \Lambda$, such that the dark quark mass may be neglected even for large $N$.

In the following we review (well-known) large-$N$ results for the spectrum and interactions of the dark bound states,
proceeding to derive estimates of the relevant interaction cross-sections that control the thermal history of this secluded sector.
We refer the reader to Refs.~\cite{Coleman:1985rnk, Manohar:1998xv,Tong:2018gt} for an extensive review of these large-$N$ results and associated topological arguments.
Of particular importance is the phenomenology of baryons in the large-$N$ limit, first discussed by Witten in Ref.~\cite{Witten:1979kh}.

\subsection{Spectrum and large-$N$ scaling}
\label{sec:spec}
Under confinement, the spectrum of the theory consists of mesons, baryons, as well as glueballs, 
in the usual array of ground states plus excited resonances.
With only one dark quark flavor, the spectrum of the confined theory features a single pseudo-Nambu-Goldstone Boson (pNGB) meson, 
associated with the breaking of the accidental axial $U(1)$ symmetry: 
The analog of the Standard Model (SM) $\eta'$, which we denote as $\etap$. 
In the absence of a dark photon or leptons, the $\etap$ is the lowest lying state, and therefore accidentally stable. 
However, $\etap$ number is not protected by any accidental symmetry, permitting number-changing interactions.
The spectrum further contains a single ground state baryon $\deltaN \sim (q)^N$ that is the analog of the SM $\Delta^{++}$.
The $\deltaN$ must be a spin-$\frac{N}{2}$ state and carries charge $\propto N$ under the accidental vector $U(1)$ -- baryon number -- of the dark sector.
(Gravitational interactions of higher-spin particles such as these are thought to be subject to causality constraints, 
which may be resolved via glueball or other gravity sector interactions~\cite{Kaplan:2019soo,Afkhami-Jeddi:2018apj,Kaplan:2020tdz}.)

The leading-order large-$N$ scaling of meson or baryon correlators is fully characterized by well-known topological arguments
that describe the scaling of the underlying correlators of quark or gluon operators.
Of particular importance is the $\etap$ decay constant $f_\etap$, defined via  $\langle 0 |J_5^{\mu} | \etap \rangle =  f_\etap p^\mu$, 
with $J_5^{\mu} = \bar{q} \gamma^\mu \gamma^5 q$. Combined with na\"\i ve dimensional analysis (NDA) arguments~\cite{MANOHAR1984189,Georgi:1986kr, Georgi:1992dw}, 
the decay constant scales as
\begin{equation}
	\label{eqn:fetap}
	f_\etap \sim \frac{\sqrt{N} \Lambda}{4\pi}\,.
\end{equation}
Similarly, the correlator $\langle 0 | G \widetilde{G} | \etap \rangle \sim \sqrt{N}$. 
Combined with the axial $U(1)$ anomaly $\partial_\mu J_5^{\mu} \sim g^2/(16\pi^2 N) G\widetilde{G}$, 
this immediately implies that the $\etap$ mass scales as~\cite{Veneziano:1976wm,WITTEN1979269}
\begin{equation}
	\label{eqn:etam}
	m_\etap^2 \sim \frac{\Lambda^2}{N}\,,
\end{equation}
in which we have assumed that additional contributions $\sim m_q \Lambda$ are always comparably negligible, for any finite $N$ we consider. 
For $N$ large, the $\etap$ is then parametrically light compared to the confinement energy scale $\sim \Lambda$.

Since $\Lambda$ characterizes the typical kinetic energy scale of light degrees of freedom inside a condensate, the baryon mass~\cite{Witten:1979kh} 
\begin{equation}
	\label{eqn:deltam}
	m_\deltaN \sim N \Lambda\,.
\end{equation} 
(This also follows from $N$-scaling of intrabaryon many-body interactions, or, when viewed as skyrmions, the baryon mass follows from the expected scaling of the mass proportional to the inverse coupling).
Similarly, one expects the lightest vector meson, the $\omega_{\text{d}}$ to have mass $\sim \Lambda$, 
and the lightest glueball, $G_d$, with $J^{PC} = 0^{++}$, to have a mass $\gtrsim \text{few}\times\Lambda$ (see e.g. Ref.~\cite{Forestell:2016qhc}).
These may decay to the $\etap$ via $\omega_{\text{d}} \to 3\etap$ and $G_d \to 2\etap$, respectively,  
with amplitudes $\sim 1/N$, such that their lifetimes $\sim N^2/\Lambda$.
The spectrum is summarized in Table~\ref{tab:confspec}.

The typical mass-splitting of excited versus ground states is expected to be $\sim \Lambda \gg m_\etap$.
Hence, we expect all excited states to decay strongly to combinations of $\etap$ and $\deltaN$, 
as allowed by parity and angular momentum conservation, but subject to suppression by powers of $1/\sqrt{N}$:
I.e., an excited meson state decay to $(\etap)^p$ has amplitude $\sim N^{(1-p)/2}$; an excited baryon decay to $ (\etap)^p \deltaN$ has amplitude $\sim N^{(2-p)/2}$;
an excited glueball decay to $(\etap)^p$ has amplitude $\sim N^{-p/2}$. 
Thus, typically the longest-lived ground or excited state has a lifetime $\lesssim N^2/\Lambda$ or $\lesssim N^3/\Lambda$.\footnote{
The values of $N$ and $\Lambda$ we consider easily satisfy $N^3 \lll M_{\text{pl}}/\Lambda$, such that the ground or excited state decay rates nonetheless always remain cosmologically efficient, 
and these states therefore have a negligible effect on the cosmological evolution of the dark sector.}

We emphasize that the mass relations~\eqref{eqn:etam} or~\eqref{eqn:deltam} are only scalings, and should typically contain $\mathcal{O}(1)$ prefactors.
However, for the sake of a concrete benchmark, in our numerical analyses below we shall treat all such prefactors as nuisance parameters, and set them all to unity.
I.e., we take $m_\etap = \Lambda/\sqrt{N}$ and $m_\deltaN = N \Lambda$,
keeping in mind that we expect our results will be relatively insensitive at the qualitative level to $\mathcal{O}(1)$ variation in these nuisance parameters.

\newcolumntype{Y}{ >{\centering\arraybackslash $} m{2cm} <{$} }
\newcolumntype{W}{ >{\centering\arraybackslash $} m{3cm} <{$} }
\begin{table}[t]
	\renewcommand{\arraystretch}{1.2}
	\centering
	\begin{tabular}{@{\extracolsep{\fill}}WYYY}
		\hline\hline
		\text{State} & \text{Mass}  & \text{Lifetime} & U(1)_V\\
		\hline
		\etap & \sim \Lambda/\sqrt{N}  & \text{stable} & 0 \\
		\deltaN & \sim N \Lambda & \text{stable} & N\\
		\hline
		\omega_{\text{d}} & \sim \Lambda & \sim N^2/\Lambda & 0 \\
		G_{\text{d}} & \sim \text{few} \times \Lambda & \sim N^2/\Lambda & 0 \\
		\hline\hline
	\end{tabular}
	\caption{Spectrum of lowest-lying bound states in the confined theory.}
	\label{tab:confspec}
\end{table}

\subsection{$\etap$ interactions}
\label{sec:etapint}
Just as in multiflavor theories, the dynamics of the $\etap$ may be represented by a chiral Lagrangian, with the chiral field $\Sigma = e^{i \etap/f_\etap}$. 
With just one quark flavor, however, the chiral theory becomes trivial, because e.g. $\Sigma^\dagger \partial_\mu \Sigma  = i\partial_\mu\etap/f_\etap$, 
so that the kinetic term is simply $\partial_\mu \etap \partial_\mu \etap$. 
Neglecting chiral symmetry breaking terms from the negligible quark masses, 
higher point $\etap$ interactions then only arise through higher-order derivative interactions -- in turn stemming from terms $\sim f_\eta^2/\Lambda^{n-2} (\Sigma^\dagger \partial_\mu \Sigma)^n$  --  
that are suppressed by powers of the chiral symmetry breaking scale $\Lambda$. 

Large-$N$ scaling and normalization arguments imply that $n$-point $\etap$ interaction amplitudes must scale as $N^{1-n/2}$.
Combined with NDA arguments, one can immediately write down the general form of the $\etap$ Lagrangian,
\begin{equation}
	\label{eqn:Letap}
	\mathcal{L}_\etap = \frac{1}{2}\partial_\mu \etap \partial_\mu \etap + \frac{m_\etap^2}{2}  \etap\etap 
	+ \sum_{k = 1} \frac{\lambda_k}{n_k!} \bigg[\frac{16 \pi^2}{\Lambda^{4} N}\bigg]^k(\partial \etap \cdot \partial \etap)^{k+1}\,, 
\end{equation}
in which $n_k = 2(k+1)$. The presence of the $n_k!$ factor follows from the expectation that the derivative expansion should remain perturbative once combinatoric factors are included.
Alternatively, one may begin with the chiral Lagrangian and apply Eq.~\eqref{eqn:fetap}, from which it follows that $\lambda_k$ are expected to be $\mathcal{O}(1)$ numbers.
(From a purely effective field theory perspective, perturbative UV completions exist in which $\lambda_k$ can take arbitrarily large (or small) values in a technically natural way.)
For any given $n$-point interaction, loop-level contributions arising from higher-order operators enter at higher order in $1/N$, 
such that in the large-$N$ limit it is sufficient to consider only tree-level $\etap$ interactions in Eq.~\eqref{eqn:Letap}.
Parity conservation requires that only even $n$-point interactions arise, with $\etap$ number changes by multiples of $2$, as in the $\etap$ Lagrangian~\eqref{eqn:Letap}.\footnote{
While the leading order number-changing interaction is then the $4\etap \to 2\etap$ process with amplitude $\sim 1/N^2$, 
one also may consider a number-changing interaction such as $3\etap \deltaN \to \etap \deltaN$, which scales as $1/N$ (cf. Sec.~\ref{sec:deltascat}), which is lower-order.
However, the thermally-averaged cross-section will be heavily suppressed by the (typically out-of-equilibrium) $\deltaN$ number density, 
which is itself bounded above by the Boltzmann-suppressed equilibrium density $\sim e^{-N}$. Thus these interactions may be neglected.
} 

In the thermal history of this dark $SU(N)$ sector (see Sec.~\ref{sec:thermal}), number-changing interactions of the $\etap$ exponentially slow the cooling of the $\etap$ plasma.
At tree-level, the amplitude for the number-changing interaction $n_i \etap \to (n- n_i) \etap$ scales as
\begin{equation}
	\label{eqn:ampn}
	A_{n_i \to n - n_i} \sim \bigg[\frac{4\pi}{\Lambda^2\sqrt{N}}\bigg]^{n - 2} (p\ccdot p)^{n/2} \sum_{j} c_{n;j} \prod_{k =1}^{n/2-1} \bigg(\frac{\lambda_k}{n_k!}\bigg)^{a^n_{k;j}}\,,
\end{equation}
in which $p$ is the typical momentum scale of the external states,
$a^n_{k;j} \in \mathbb{N}$ belongs to the $j$th solution of the Diophantine equation 
\begin{equation}
	\sum_{k=1}^{n/2-1} 2k\,a^n_k = n -2\,,
\end{equation}
and $c_{n;j}$ is a combinatoric factor.
For instance, for the $2 \to 2$ amplitude ($n=4$) the unique solution is simply $a^4_1 = 1$; 
for the $4 \to 2$ amplitude ($n=6$) there are two solutions $a^6_1 = 2$, $a^6_2 = 0$ and $a^6_1 = 0$, $a^6_2 = 1$.
The total number of vertices for the $j$th solution $v_{n;j} = \sum a^n_{k;j}$. 
The factor $c_{n;j}$ contains the particle permutations $n!$, along with the usual symmetry factor and internal lines permutation factor, such that, schematically
\begin{equation}
	\label{eqn:cnj}
	c_{n;j} \sim n! \frac{v_{n;j}!}{\prod_k a_k!} \prod_{\text{vertices}} {}^{n_k}\!P_{\text{lines}}\,.
\end{equation}

When combined with the $\sim 1/n!$ symmetry factors for the phase space integral, 
Eqs.~\eqref{eqn:ampn} and~\eqref{eqn:cnj} taken together imply that the $n$-point scattering cross-section scales as $n!/[N^n (\prod_k n_k!)^2] \sim 1/(N^n n!)$, 
so that the lowest $n$-point interactions dominate.
Thus for our purposes, we need only consider the $2 \to 2$ and $4 \to 2$ processes, whose amplitudes scale as
\begin{align}
	A_{2\etap \to 2\etap} & \sim \bigg[\frac{4\pi}{\Lambda^2\sqrt{N}}\bigg]^2 (p\ccdot p)^{2} \lambda_1\,,\nonumber\\
	A_{4\etap \to 2\etap} & \sim \bigg[\frac{4\pi}{\Lambda^2\sqrt{N}}\bigg]^4 (p \ccdot p)^3 \Big( 10 \lambda_1^2 + \lambda_2 \Big)\,,
\end{align} 
where the factor of $10$ enters as $4^2/2! \times 6!/(4!)^2$.
With respect to the invariant mass of the incoming $\etap$ pair $s$, in the $\sqrt{s} \gg \Lambda$ regime, the corresponding $2 \to 2$ and $2 \to 4$ cross-sections are then estimated as
\begin{equation}
	\label{eqn:eta2242}
	\sigma_{2\etap \to 2\etap}(s) \sim \frac{\pi^3 s^3 |\lambda_1|^2}{4\Lambda^8 N^2}\,,\qquad \sigma_{2\etap \to 4\etap}(s) \sim \frac{\pi^3 s^7}{48\Lambda^{16} N^4}\big| 10 \lambda_1^2 + \lambda_2 \big|^2\,.
\end{equation}
The overall normalization of the full tree-level cross-section arising from Eq.~\eqref{eqn:Letap} is expected to include additional numerical factors from the full phase space integral 
(see Sec.~\ref{sec:numxsec} below), that is only roughly estimated here.

The thermally-averaged cross-section of the $2 \to 4$ process will be important in determining the $\etap$ freeze-out and relic abundance.
Defining the temperature of the dark sector to be $T_d$, and $x_d \equiv m_{\etap}/T_d$, then expanding to leading order in the non-relativistic regime $1/x_d \ll 1$
\begin{align}
	\langle \sigma v \rangle_{2 \to 4}  
	& = \frac{x_d}{8 m_{\etap}^5 \big(K_2(x_d)\big)^2}\int_{16m_{\etap}^2}^\infty ds \sqrt{s}(s - 4 m_{\etap}^2)K_1\big[\sqrt(s)/T_d\big] \sigma_{2\etap \to 4\etap}(s) \label{eqn:sigmavetap} \\
	& \simeq \frac{ \zeta\, x_d^{1/2} \, e^{-2x_d}  }{N^{11} \Lambda^2}\,, \qquad \zeta \simeq 10^{-2}\big|10\lambda_1^2 + \lambda_2\big|^2\,, \nonumber
\end{align}
in which $K_n$ is the $n$th modified Bessel function of the second kind, and $\zeta$ contains a numerical prefactor, 
whose value anticipates the result of the numerical treatment following in Sec.~\ref{sec:numxsec}.
Here we have enforced the $2 \to 4$ phase space kinematic constraint $s \ge 16m_\etap^2$ in the integration limit.
This estimate will inform our expectations of the behavior of the $\etap$ freeze out, discussed below.
However, we emphasize that for our numerical studies we will use the full expression derived from Eq.~\eqref{eqn:Letap} for all regimes of $x_d$, 
with the thermal average performed by numerical integration.

\subsection{Numerical $\etap$ cross-sections}
\label{sec:numxsec}

\begin{figure}[t]
    \centering
    \begin{subfigure}[b]{0.45\textwidth}
        \centering
        \includegraphics[scale=0.38]{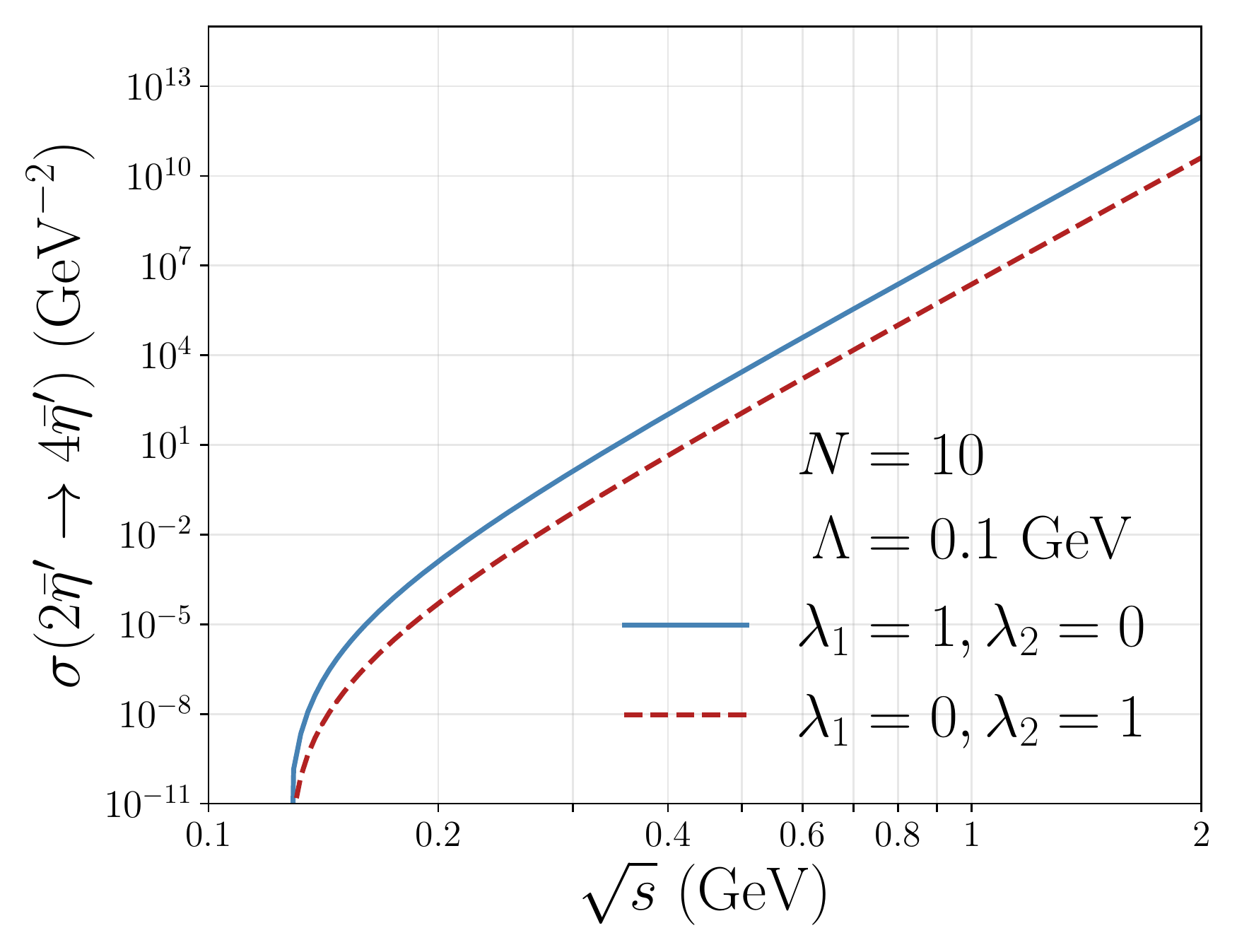}
        \caption{}
        \label{fig:eta_cs}
    \end{subfigure}\hfill
    \begin{subfigure}[b]{0.45\textwidth}
        \centering
        \includegraphics[scale=0.38]{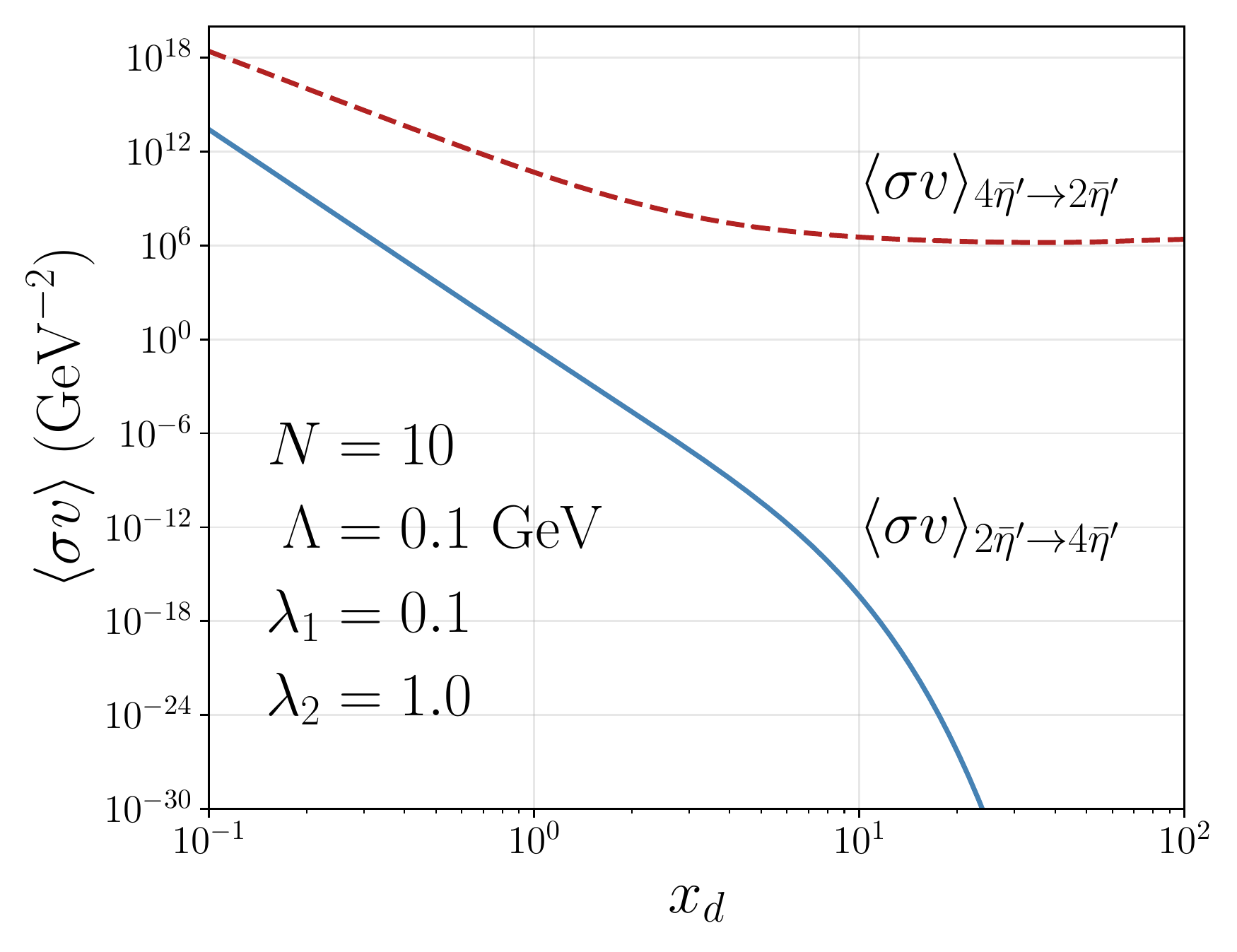}
        \caption{}
        \label{fig:eta_tcs}
    \end{subfigure}
    \caption{Left: Contributions from the 4-pt and 6-pt terms of the chiral Lagrangian to the zero-temperature cross section for $2\etap\to4\etap$. Right: Rescaled thermally-averaged cross section for $2\etap\leftrightarrow4\etap$ with $\lambda_1 = 0.1$ and $\lambda_2=1$.}
    \label{fig:eta_zt_cs_tcs}
\end{figure}

In order to compute the full cross section for $2\etap\to4\etap$ we first use \texttt{FeynRules v2.3}~\cite{christensen2009feynrules} to generate model files for \texttt{FeynArts}~\cite{hahn2001generating}. 
We then compute the full matrix elements using \texttt{FeynArts v3.11} and \texttt{FeynCalc v9.3}~\cite{shtabovenko2016new}. 
To perform integration over phase space, we use the simple Monte Carlo phase-space generator \texttt{RAMBO}~\cite{kleiss1985new}. 
We verify our results using \texttt{MadGraph5 v2.7}~\cite{alwall2011madgraph}. 
Fig.~(\ref{fig:eta_cs}) shows the $2\etap\to4\etap$ cross section for two choices of $\lambda_1$ and $\lambda_2$, with $N= 10$ and $\Lambda = 0.1$\,GeV, and taking $m_{\etap} = \Lambda/\sqrt{N}$.
While the scaling with respect to $N$, $\Lambda$ and $s$ of the numerical $2\to 4$ cross-section matches that of Eq.~\eqref{eqn:eta2242} in the $\sqrt{s} \gg \Lambda$ regime, 
the overall normalization of the numerical result is a factor of $\sim 10^{-6}$ smaller compared to our estimates in Sec~\ref{sec:etapint}.
This is likely due to a number of numerical factors arising from the phase space integral, that was only estimated above.

With the numerical cross section for $2\etap\to4\etap$ in hand, we use Eq.~\eqref{eqn:sigmavetap} to compute the thermally-averaged cross section $\expval{\sigma v}_{2\etap\to4\etap}$, 
and derive the $4\etap\to2\etap$ via detailed balance, i.e., $\expval{\sigma v}_{2\etap\to4\etap}n^2_{\etap,\text{eq}} = \expval{\sigma v}_{4\etap\to2\etap}n^4_{\etap,\text{eq}}$.
Fig.~\ref{fig:eta_tcs} shows the thermally-averaged cross sections for the benchmark choice $\lambda_1=0.1$ and $\lambda_2=1$, with $N= 10$ and $\Lambda = 0.1$\,GeV.

\subsection{$\deltaN$ scattering} 
\label{sec:deltascat}
Pair production of the $\deltaN$ baryons from $\etap$ annihilation -- i.e. $\etap\etap \to  \deltaN \deltaNb$ -- occurs at no order in the $1/N$ expansion, and instead, 
by analogy to e.g. monopole pair production, proceeds via an exponentially suppressed amplitude $\sim e^{-N}$~\cite{Witten:1979kh}. 
This can also be understood as a consequence of a combinatoric argument: If the probability to produce (annihilate) a single color quark pair $\sim w$, 
then one would expect the probability to produce (annihilate) the $N$ colors for the baryon bound state to scale as $w^N = \exp[-|\log(w)| N]$.
That is, the 4-point amplitude
\begin{equation}
	A_{\etap\etap \to  \deltaN \deltaNb} \sim e^{-c N}\,, \qquad c > 0\,.
\end{equation}
This exponential scaling is crucial to the identification of the $\deltaN$ as a possible freeze-in dark matter candidate.
The value of the factor $c$ is not known \emph{a priori}. (The amplitude for the crossed process $\etap \deltaN \to \etap \deltaN$ scales as $\sim 1$ in the large $N$ limit. 
However, the Hartree approximation approach, in which each quark is treated as a independent degree of freedom in the potential of the $N-1$ others, 
suggests that the baryon in this process acts only as a background field and is undeflected. 
Rather, the amplitude for $\etap \deltaN \to \etap \deltaN$ at \emph{fixed velocity} change for the baryon is similarly exponentially suppressed. 
This can be understood as a consequence of wavefunction overlap of the interacting quark with each spectator, raised to power $N$.)
At the $\deltaN$ threshold, 
the corresponding cross-section
\begin{equation}
	\label{eqn:etadeltaxsec}
	\sigma _{\etap\etap \to  \deltaN \deltaNb} \sim \frac{e^{- 2c N}}{64\pi N^2 \Lambda^2}\,.
\end{equation}
The corresponding thermally-averaged cross-section, defined with respect to the $x_d \equiv m_\etap/T_d$
\begin{equation}
	\label{eqn:svdelta}
	\langle \sigma v \rangle_{\etap\etap \to  \deltaN \deltaNb} \sim \frac{N^{13/4} (x_d)^{1/2} e^{-2cN} e^{-2(N^{3/2}-1)x_d}}{32 \pi^{3/2} \Lambda^2}\,.
\end{equation}
at leading order in the $\etap$ non-relativistic regime $1/x_d \ll 1$. 

Finally, the $\deltaN$ 4-point self-interaction scales $\sim N$ in the large-$N$ limit,
which also follows from the expected general 4-Fermi operator form $\sim \deltaNb \deltaN \deltaNb \deltaN/f_\deltaN^2$ from NDA arguments, 
with $f_\deltaN$ the $\deltaN$ decay constant obeying a similar relation as Eq.~\eqref{eqn:fetap}.
We will only be interested in the limit that the $\deltaN$ baryons are non-relativistic, so that the $\deltaN \deltaN \to \deltaN \deltaN$ cross-section 
\begin{equation}
	\label{eqn:delta22}
	\sigma_{2\deltaN \to  2\deltaN} \sim \frac{4\pi^3}{\Lambda^2}\,,
\end{equation}
under NDA. Unlike processes involving the $\etap$, this process does not vanish in the large-$N$ limit. A similar result applies to $\deltaN \deltaNb \to \deltaN \deltaNb$ scattering.

\section{Dark cosmology}
\label{sec:darkcos}

\subsection{Thermal baths}
\label{sec:thermal}
Because the secluded dark sector is fully decoupled from the SM, the cosmological setup involves two thermal baths:
A dark bath with temperature $T_d$; and an SM bath with temperature $T_{\SM}$. 
In this discussion we shall generally always assume that the energy budget is dominated by the SM sector, so that
\begin{equation}
	\label{eqn:SMdom}
	g_{*,\SM} \gtrsim g_{*,d} (T_d/T_{\SM})^4\,,
\end{equation}
where $g_*$ is as usual the effective number of relativistic degrees of freedom. 
As a consequence, the equation of state and Hubble scale evolution is dominated by the SM sector, 
and in this discussion one can then approximate $H \simeq (8\pi^3/90)^{1/2}\, g_{*,\SM}^{1/2} T_{\SM}^2/M_{\text{pl}}$.
Note that in the numerical results presented below, \emph{all terms} in the energy density are retained, and we characterize the parametric range for which SM domination remains.
The assumption of SM domination may be relaxed for the case of pure $\deltaN$ dark matter freezing in the from the $\etap$ plasma, considered in Sec.~\ref{sec:deltaDM} below
(see Ref.~\cite{Erickcek:2020wzd} for a discussion of the effects of an early cannibal dominated era).
 
At early times, before confinement, the dark sector is composed of a thermal plasma of dark gluons and quarks, with number of degrees of freedom $g_d = 2(N^2-1) + (7/8)\,4N$.
Once this plasma reaches a temperature $T_{d,c} \lesssim \Lambda$, the sector undergoes confinement. 
As discussed in Sec.~\ref{sec:spec}, 
this produces a spectrum of bound states that includes not only a (semi)relativistic population of the lowest lying $\etap$ state -- $m_{\etap}/T_{d,c} \sim 1/\sqrt{N} \ll 1$ -- 
but also heavier meson and glueball states with masses $\gtrsim \Lambda$, as well as the $\deltaN$ and baryon excited states with masses $ \gtrsim N \Lambda \gg T_{d,c}$.
We typically consider $\Lambda \ll M_{\text{pl}}$ and $N \lesssim 10^2$.
The excited state lifetimes $\tau \sim N^k/\Lambda$, with typically $k$ a small positive integer, satisfy $\tau \ll 1/H$ for these typical values.
Hence all excited states swiftly decay to $\etap$ and/or $\deltaN$ final states.
In addition, the $\deltaN$ production is nominally heavily Boltzmann suppressed were they to reach equilibrium abundances. 
As a consequence, we can imagine the post-confinement dark sector to comprise predominantly a plasma of $\etap$ mesons in local thermodynamic equilibrium, with temperature $T_d$,
along with an (at most) exponentially small population of $\deltaN$ baryons: We shall discuss the latter further below.

While the dark and SM sector each remain in local thermodynamic equilibrium, the ratio of the entropy densities is a conserved quantity of the scale evolution. 
That is, the entropy ratio
\begin{equation}
	\label{eqn:entropyratio}
	\rs \equiv \frac{s_d}{s_{\SM}}  \simeq \frac{45 x_\SM^3}{4\pi^4 g_{*,\SM}}x_d^{-3} \int_{x_d}^{\infty} dy \frac{(2y^2 -x_d^2)\sqrt{y^2 -x_d^2}}{\exp(-y) -1}\,,
\end{equation}
in which we have assumed the $\etap$ dominates the dark entropy, and defined the mass-temperature ratio parameters
\begin{equation}
	x_{\SM} =  m_{\etap}/T_{\SM}\,, \qquad \text{and} \qquad x_d \equiv m_{\etap}/T_d\,. 
\end{equation}
In particular, if $g_{*,\SM}$ has only subleading dependence on the SM sector temperature in a particular epoch, then Eq.~\eqref{eqn:entropyratio} permits $x_{SM}$ to be written explicitly in terms of $x_d$, viz.
\begin{equation} 
	x_\SM  \simeq \bigg[\frac{45}{\rs 4\pi^4 g_{*,\SM}} x_d^{-3} \int_{x_d}^{\infty} dy \frac{(2y^2 -x_d^2)\sqrt{y^2 -x_d^2}}{\exp(-y) -1}\bigg]^{-1/3}\,.
\end{equation}
As the dark sector cools, and while equilibrium is maintained, the $\etap$ become non-relativistic such that the dark sector energy density $s_d \simeq x_{d} n_{\etap, \eq}$.
That is, the entropy ratio simplifies to
\begin{equation}
	\rs \simeq \frac{45}{2^{5/2}\pi^{7/2} g_{*,\SM}} x_{\SM}^3 x_d^{-1/2}e^{-x_d}\,, \label{eqn:entropyratioNR}
\end{equation}
so that $x_{\SM} \sim \rs x_d^{1/2}e^{x_d}$. 
The dark sector can thus cool exponentially slower than the SM. 
If the $\etap$ are instead relativistic then simply
\begin{equation}
	\label{eqn:entropyratioR}
	\rs \simeq T_d^3/ \big(g_{*,\SM} T_{\SM}^3\big)\,,
\end{equation}
so that the temperatures redshift together.

The relation~\eqref{eqn:entropyratioNR} can be further re-expressed in terms of the temperature ratio
\begin{equation}
	\label{eqn:xirelation}
	\xi \equiv \frac{T_d}{T_{\SM}} \simeq \Big[ (2^{5/2}\pi^{7/2} g_{*,\SM}/45) \rs x_d^{-5/2}e^{x_d}\Big]^{1/3}\,.
\end{equation}
E.g. for a process that freezes out at $x_{d,f} \sim 10$, such as the $\etap$ number-changing $4 \to 2$ process (see Sec.~\ref{sec:etadeltaDM}), then $\xi_f \simeq 7 (g_{*,SM} \rs)^{1/3}$.
Requiring $\rs \lesssim \text{few} \times 10^{-5}$ then ensures the SM domination condition~\eqref{eqn:SMdom} is satisfied even for semi-relativistic $\etap$ at freeze-out 
(in practice, a larger ratio can be tolerated as the $\etap$ typically freeze out non-relativistically).
Assuming $g_{*,\SM} \sim g_{*,d}$ at early times, this corresponds to an asymptotic temperature ratio $\xiinf\lesssim \text{few} \times 10^{-2}$,
where
\begin{equation}
	\xiinf \equiv T_{d,\infty}/T_{\SM,\infty}\,,
\end{equation}
is the temperature ratio of the dark sector to the SM sector at asymptotically early times, well before both the electroweak and dark confinement phase transitions.
Such a large entropy or temperature ratio can arise in a variety of scenarios, for example an early out-of-equilibrium decay from a matter-dominated phase predominantly into the SM.
This can arise because of Bose enhancements for decays to the SM Higgs, noting no such scalar is present in the dark sector (see e.g. Ref~\cite{Adshead:2019uwj}).

Hereafter, we will characterize the dark sector thermal history with respect to $\xiinf$, rather than the entropy ratio $r_s$. 
Assuming that the confining phase transition of the dark sector generates little entropy,
\begin{equation}
	\label{eqn:entcons}
 	r_s \equiv \frac{h_d(T_d)}{h_\SM(T_\SM)}\xi^3 \simeq \frac{h_{d,\infty}}{h_{\SM,\infty}}\xiinf^3\,,
\end{equation}
where $h_d$ ($h_{\SM}$) denotes the effective number of entropic degrees of freedom stored in the dark (SM) sector.
Since $g_d \sim 2 N^2$ for the deconfined sector in the large-$N$ limit, then for a fixed $\xiinf$ the entropy ratio itself scales as 
\begin{equation}
	\label{eqn:rsxiinf}
 	r_s \simeq \frac{2N^2}{g_{*,\SM,\infty}}\xiinf^3\,,
\end{equation}
and we take $g_{*,\SM,\infty} \simeq 10^2$.

\begin{figure}[t]
    \centering
    \includegraphics[width=10cm]{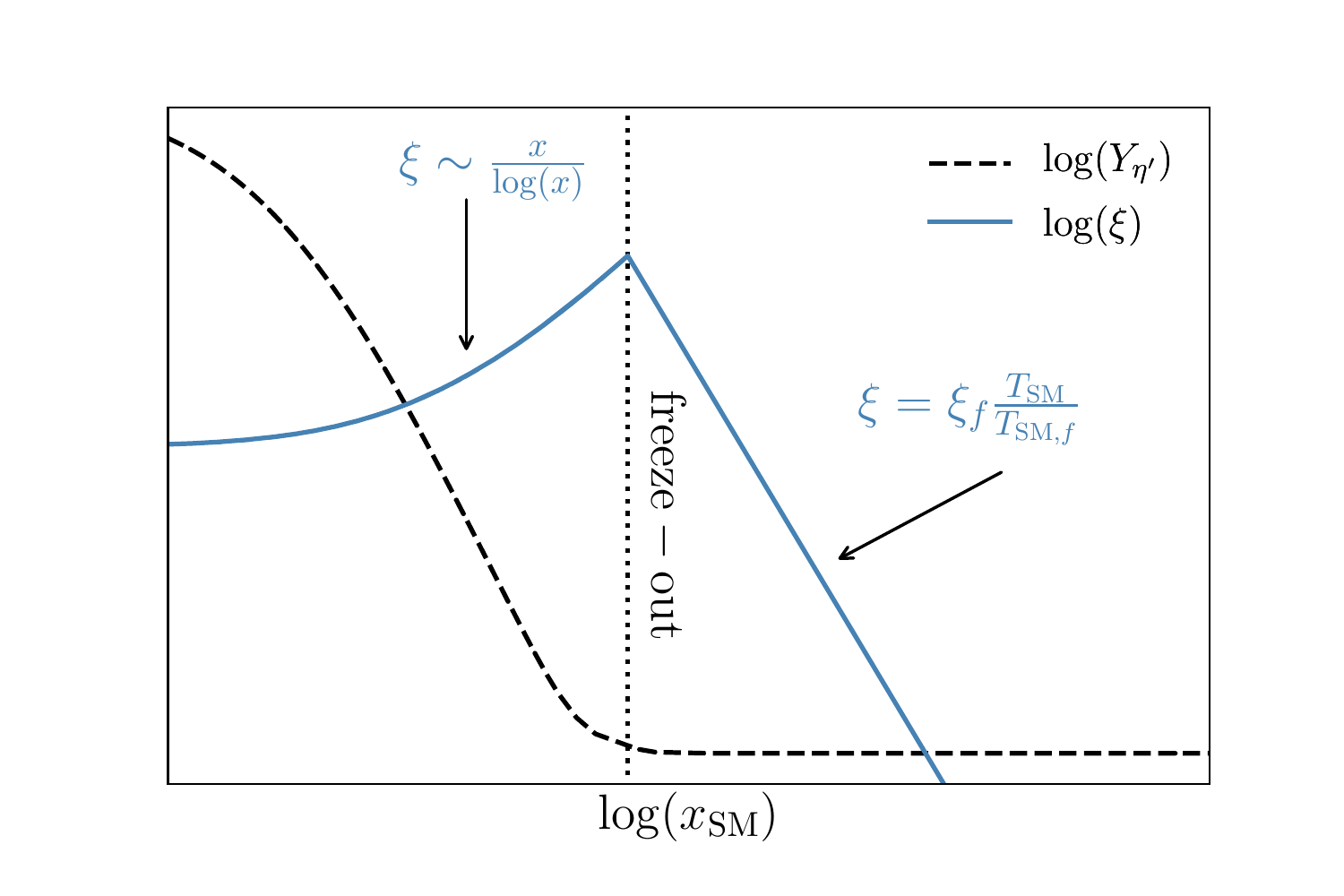}
    \caption{Schematic plot of the behavior of $\xi \equiv T_{\rm d}/T_{\rm SM}$. For $\etap$ non-relativistic prior to freeze-out, $\xi$ increases like $x_\SM/\log(x_\SM)$. 
    After the $\etap$ has frozen out, $\xi$ simply redshifts quadratically. }
    \label{fig:xi_vs_x}
\end{figure}

The dark sector eventually leaves chemical and kinetic equilibrium at a `freeze-out' temperature $T_{d,f}$, i.e. the $\etap$ self-decouple. 
Assuming the energy-dominant SM sector is radiation-dominated, the dark sector temperature subsequently redshifts as
\begin{equation}
	\label{eqn:Tdredshift}
	T_d = T_{\SM}^2 \frac{T_{d,f}}{T_{\SM,f}^2}\,,
\end{equation}
cooling quadratically faster with respect to the SM sector. The dark thermal history is shown schematically in Fig.~\ref{fig:xi_vs_x}.

\subsection{Constraints}
\label{sec:cons}
Before turning to discuss production and abundances of the $\deltaN$ and $\etap$, we briefly anticipate two important astrophysical/cosmological constraints on the dark cosmology.

The first constraint arises via DM self-interaction bounds, and is relevant to both $\deltaN$ and $\etap$ DM species.
The best limits on dark matter self-interaction are generated by detailed fits to DM halos of galaxies and clusters~\cite{Kaplinghat_2016}. 
This gives an upper limit,
\begin{equation}
	\label{eqn:selfint}
	\sigma_{\text{SI}} / m \lesssim 0.1 \ \text{cm}^2/\text{g} \sim 0.1 \text{b}/\text{GeV}\,. 
\end{equation}
For the case that the DM is either purely made of $\etap$ or of $\deltaN$, 
this constraint may be applied straightforwardly via the $\etap$ and $\deltaN$ self-interaction cross-sections in Eqs.~\eqref{eqn:eta2242} and~\eqref{eqn:delta22}, respectively.
Self-interaction constraints for the case in which both species significantly contribute to DM -- two-component DM -- require a dedicated analysis of the relevant observational data.
However, the two-component self-interaction constraints are bounded above by those for pure $\etap$ and $\deltaN$ DM.

The second constraint, relevant mainly for the $\etap$, are bounds on the effective number of relativistic degrees of freedom, $N_{\text{eff}}$, 
at the Big Bang Nucleosynthesis (BBN) and Cosmic Microwave Background (CMB) epochs. 
Measurements from 2018 Plank and BBN set $N_{\text{eff}}$ to be~\cite{Cyburt_2016,PhysRevD.98.030001}
\begin{align}
    N^{\text{CMB}}_{\text{eff}} &= 2.92 \pm 0.36 && (\text{TT,\ TE,\ EE+lowE})\\
    N^{\text{BBN}}_{\text{eff}} &= 2.85 \pm 0.28 && (\text{BBN}+Y_{P}+D)\,.
\end{align}
After self-decoupling, we treat the $\etap$ as a decoupled non-relativistic population, having an equilibrium phase space density but at a redshifted temperature as in Eq.~\eqref{eqn:Tdredshift}.
Assuming the $\etap$ decoupling occurs before BBN and CMB epochs, their contribution 
\begin{equation}
	\delta N_{\text{eff}}(T) = \frac{60\sqrt{2}}{7\pi^{7/2}} \bigg[\frac{m_{\etap}}{T_{\nu}}\bigg]^4 \frac{e^{-x_d(T)}}{x_d(T)^{3/2}}\,, \qquad x_d(T) = \frac{m_{\etap}^2}{T^2} \frac{1}{x_{d,f}\xi^2_{f}}\,,
\end{equation}
in which $T_{\nu}$ is the neutrino temperature, $x_{d,f}$ is the freeze-out dark parameter
and the corresponding freeze-out ratio $\xi_{f}$ is determined from $x_{d,f}$ via the entropy ratio relation~\eqref{eqn:xirelation} and~\eqref{eqn:rsxiinf}.
Revisiting the above case of a freeze-out at $x_{d,f} \sim 10$, then $x_d(T) \sim 10^{-2} (m_{\etap}^2/T^2) (N^2 g_{*,\SM,f})^{-2/3} \xiinf^{-2}$.
Thus, for $m_{\etap} \sim T_{\text{BBN}} \sim $~MeV and $g_{*,\SM,f} \sim N \sim 10$, 
one conservatively requires $\xiinf\lesssim 10^{-2}$ to always ensure small contributions to $\delta N_{\text{eff}}$.
Though this bound relaxes somewhat for heavier $\etap$, 
combining this discussion with that of Sec.~\ref{sec:thermal} we shall take $\xiinf\lesssim 10^{-2}$ as a typical benchmark for the DM production mechanisms and numerical studies considered below.

\subsection{$\deltaN$ abundance}
\label{sec:deltaY}
The intriguing feature of the large-$N$ limit of this dark sector is the presence of the heavy $\deltaN$ baryons, with exponentially suppressed couplings to the $\etap$ plasma. 
In particular we now show the $\deltaN$ may be produced from the $\etap$ (equilibrium) plasma via a freeze-in that is insensitive to the scale $\Lambda$, while exponentially sensitive to $N$.

We assume the $\etap$ plasma remains in equilibrium throughout this freeze-in, and we further assume the $\deltaN$ abundance is always far from equilibrium, 
so that we may neglect contributions from the inverse $\deltaN \deltaNb \to \etap\etap$ process. 
Defining the SM-normalized yields $Y \equiv n/s_{\SM}$ the corresponding Boltzmann equation
\begin{align}
	\frac{dY_{\deltaN}}{d x_{\SM}} 
	& \simeq \frac{x_{\SM} s_{\SM}}{H(m_\etap)}~\langle \sigma v \rangle_{\etap\etap \to  \deltaN \deltaNb}~[Y_{\etap,\eq}]^2\,,\nonumber\\
	& \sim \frac{e^{- 2c_* N}}{64\pi N^2 \Lambda^2} \frac{x_{\SM} s_{\SM}}{H(m_\etap)}~[Y_{\etap,\eq}]^2\,. \label{eqn:expydelta}
\end{align}
Here the thermally-averaged cross-section~\eqref{eqn:svdelta} is exponentially suppressed in $N$, and we note the equilibrium yield $Y_{\etap,\eq} = \rs n_{\etap,\eq}(x_d)/s_{d}(x_d) $ 
can be written explicitly a function of the dark sector parameter $x_d$, and the entropy ratio $\rs$.
We define the initial abundance generated from confinement $Y_{\deltaN}(x_d \simeq 1/\sqrt{N}) = Y_{\deltaN}^0$. 
Since we expect similarly $Y_{\deltaN}^0 \sim e^{-2cN}$, we have absorbed this component by replacing $c$ with an effective exponent $c_*$ in Eq.~\eqref{eqn:expydelta}, 
defined to produce the same final abundance, but from the initial condition $Y_{\deltaN}^0 = 0$. 

Because the $\etap$ plasma remains in equilibrium, $x_{\SM}$ can be expressed as a function of $x_d$ via the entropy relation~\eqref{eqn:entropyratio}. 
Applying this relation generates a first-order ordinary differential equation in $x_d$.
This may be directly integrated from the confinement temperature $x_{d,c} \simeq 1/\sqrt{N}$ to $x_{d} \to \infty$. 
Some intuition for the form of the final result can be acquired by approximating the $\etap$ phase space distribution as being non-relativistic, in which case $Y_{\etap,\eq} \sim \rs/x_d$.
In this case the final abundance
\begin{equation}
	Y_{\Delta} \simeq \frac{5 \sqrt{5} N^7 \xiinf^6}{6 \pi (g_{*,\SM})^{5/2}} \frac{M_{\text{pl}}}{\Lambda} e^{-2(c_*+1)N}\,.
\end{equation}
Of crucial importance is the observation that since the $\etap$ remain in equilibrium, the final $\deltaN$ yield depends on $\Lambda$ only 
via the prefactor $M_{\text{pl}}/\Lambda$ in Eq.~\eqref{eqn:expydelta}. 
Thus the $\deltaN$ relic abundance $\Omega_{\deltaN}h^2= Y_{\Delta,\infty}m_\deltaN s_0/\rho_c$ is independent of $\Lambda$, but exponentially sensitive to $N$.

\begin{figure}[t]
    \centering
    \includegraphics[width = 10cm]{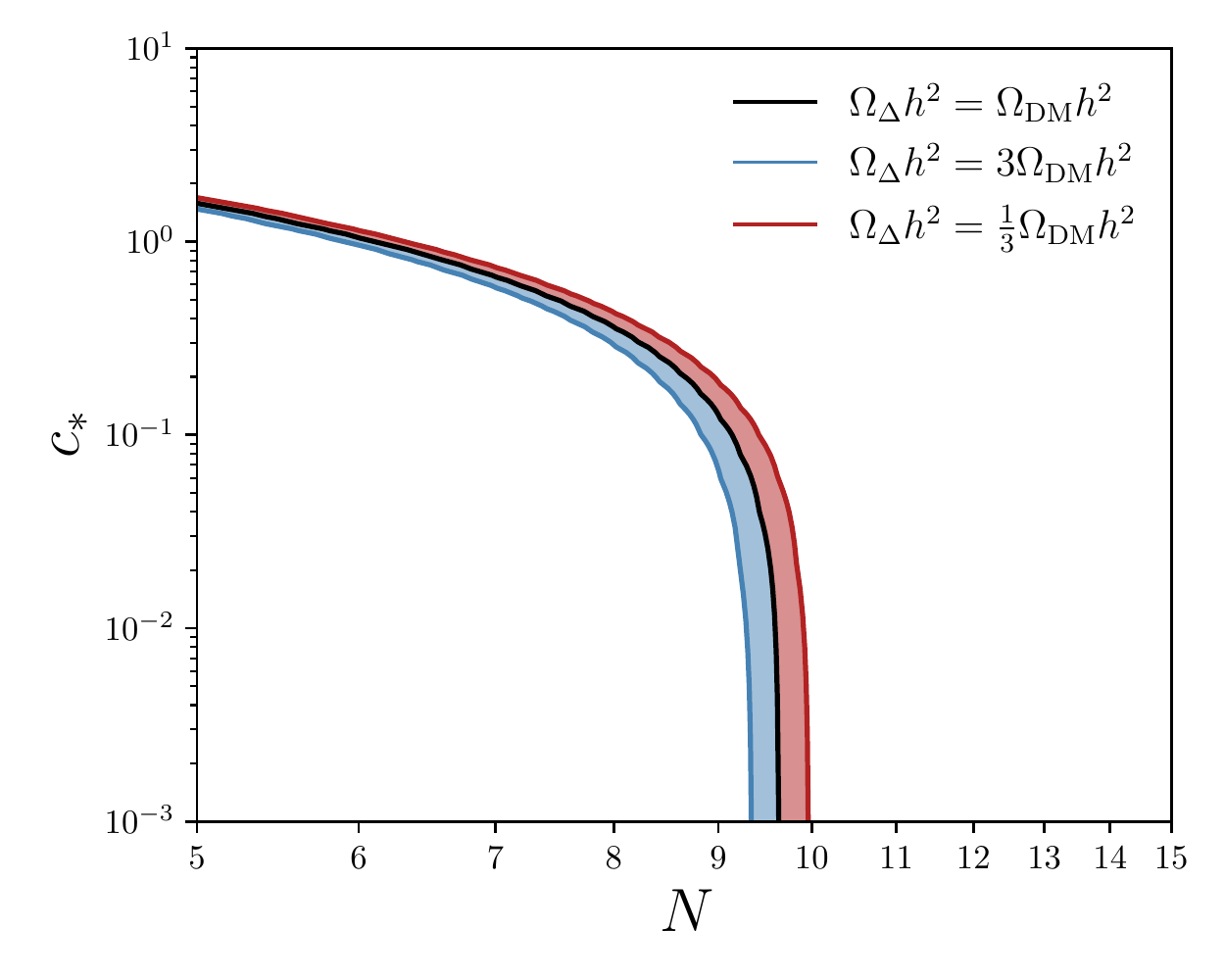}
    \caption{The contour in the $c_*$--$N$ plane, fixing $\Omega_{\deltaN} = \Omega_{\text{DM}}$. 
    Also shown is the range covering  $\Omega_{\deltaN}h^2 < 3\Omega_{\text{DM}}h^2$ (blue) and $\Omega_{\deltaN}h^2 < 1/3\,\Omega_{\text{DM}}h^2$ (red).}
    \label{fig:deltaDM}
\end{figure}

In Fig.~\ref{fig:deltaDM}, we show in the $c_*$--$N$ plane the relic-abundance contours for $\Omega_{\deltaN}h^2 = \Omega_{\text{DM}}h^2$, 
and ranges up to $3\Omega_{\text{DM}}h^2$, and down to $1/3\,\Omega_{\text{DM}}h^2$.
We compute $\Omega_{\deltaN}h^2$ by numerically integrating the coupled system given in Eq.~\eqref{eqn:boltznum} (see Sec.~\ref{sec:numstudy} for details). 
We set $\lambda_1=0.1$, $\lambda_2=1$, and $\xi_{\infty}=10^{-2}$, with the results insensitive to the choice $\Lambda=10^{-6}$\,GeV.
For $c_* \ll 1$, $\Omega_{\deltaN}h^2 \sim e^{-2 N}$ only, such that the $\deltaN$ abundance is always smaller than the DM abundance once $N \gtrsim 10$.
That is, there is a natural upper bound to the rank of the dark QCD gauge group -- i.e. $N \lesssim 10$ --  such that the $\deltaN$ can form the whole of cosmological DM.

\subsection{Pure $\deltaN$ dark matter}
\label{sec:deltaDM}

The freeze-in production of $\deltaN$ can produce a DM-like abundance.
However, one must account also for the $\etap$ plasma: Either this abundance must be small compared to the $\deltaN$, or the $\etap$ themselves must be able to annihilate or decay. 
For a moment we shall very briefly sketch out whether additional dynamics can be present in the dark sector that allows the $\etap$ to vanish, while satisfying self-interaction and other bounds.

The simplest realization of this phenomenology is to consider an additional dark photon, $\gamma_d$, from gauging the accidental vector $U(1)$ symmetry present in the dark sector.
(A discussion of a slightly more complicated multiflavor chiral dark sector featuring dark pion and/or dark baryon DM, that freeze-out from a dark photon, 
can be found in Refs.~\cite{Harigaya:2016rwr,Co:2016akw}.)
This generates the decay $\etap \to \gamma_d \gamma_d$, after which the dark photon simply redshifts as a relativistic plasma, with negligible contribution to the DM energy density in the present epoch.
Provided the $\etap$ decay by $x_d \sim 1$, the $\etap$ plasma does not significantly exponentially heat with respect to the SM. 
Typical $\delta N_{\text{eff}}$ bounds apply, via
\begin{equation}
	\label{eqn:darkphotonNeff}
	\delta N_{\text{eff}}(T) \simeq 2 \xi_f^4 \sim 10^{-2} (g_{*,\SM} N^2 \xiinf^3)^{4/3}\,,
\end{equation}
assuming $\etap$ decay while semi-relativistic and applying the corresponding approximate entropy relations~\eqref{eqn:entropyratioR} and~\eqref{eqn:rsxiinf} .

We do not apply a 't~Hooft limit to the dark photon coupling $g_V$. 
The $\etap$ width is then, via the usual chiral anomaly
\begin{equation}
	\Gamma_{\etap} = \bigg(\frac{N}{3}\bigg)^2 \frac{g_V^4m_{\eta}^3}{512 \pi^5 f_{\etap}^2} \sim \frac{g_V^4 \Lambda}{288 \pi^3 \sqrt{N}}\,.
\end{equation}
In order for the $\etap$ to live long enough to permit a freeze-in of the $\deltaN$, this width should be comparable to the Hubble scale near confinement. 
Applying the entropy relation~\eqref{eqn:entropyratioR} keeping the $\etap$ relativistic, then $H \sim g_{*,\SM}^{1/6}\rs^{-1/3} \Lambda^2/M_{\text{pl}}$. 
Hence, we require the coupling
\begin{equation}
	g_V \sim 10 N^{1/24} \xiinf^{-1/4} (\Lambda/M_{\text{pl}})^{1/4} \lesssim 1\,,
\end{equation}
the latter bound for perturbativity. For a given $N$, this amounts to an upper bound on $\Lambda$.
The dark photon also introduces a long-range self-interaction between the $\deltaN$, which scales as $g_V^2 N^2$. The cross-section
\begin{equation}
	\sigma_{2\deltaN \to  2\deltaN} \sim \frac{g_V^4 N^2}{ 64\pi \Lambda^2}\,.
\end{equation}
Combining the self-interaction bound~\eqref{eqn:selfint} with the $\Gamma_{\etap} \sim H$ constraint, this implies the bound 
\begin{equation}
 	 g_{*,\SM}^{1/6} \xiinf^{-1} N^{5/6} /\Lambda^2 M_{\text{pl}} \lesssim 1/\text{GeV}^3\,.
\end{equation}
For a given $N$, this amounts to an lower bound on $\Lambda$. 
This is typically far weaker than the bound from the $\deltaN$ self-interaction~\eqref{eqn:delta22}, that requires $\Lambda \gtrsim 1$\,GeV for $N \lesssim 10^2$.

Perturbativity is lost in the large $N$ limit. But for $N \lesssim 10^2$ and $\xiinf$ small enough to satisfy the $\delta N_{\text{eff}}$ bound from Eq.~\eqref{eqn:darkphotonNeff}, 
$\Lambda$ can be quite large -- easily avoiding the self-interaction limits -- while perturbative values of $g_V$ still exist that result in an appropriate $\etap$ lifetime.
For e.g. $N \lesssim 10^2$ and $\xiinf\sim 10^{-2}$, the $\delta N_{\text{eff}}$, perturbativity, and self-interaction bounds can be satisfied over the very large range $1 \lesssim \Lambda \lesssim 10^{12}$~GeV.

\subsection{$\etap$--$\deltaN$ dark matter}
\label{sec:etadeltaDM}
We now turn to the more complicated scenario in which the $\etap$ remains stable, and may contribute or dominate the DM relic abundance. 
The two-component $\etap$--$\deltaN$ dynamical system is described by the coupled Boltzmann equations
\begin{subequations}
\label{eqn:boltzsystem}
\begin{align}
\frac{dY_{\etap}}{d x_{\SM}} 
	 & = -\frac{dY_{\deltaN}}{d x_{\SM}} - \frac{x_{\SM} s_{\SM}}{H(m_{\etap})} \langle \sigma v \rangle_{2\etap \to 4\etap} \frac{Y^2_{\etap}}{Y^2_{\etap,\eq}}\Big[Y^2_{\etap} - Y^2_{\etap,\eq}\Big] \,,\label{eqn:boltzeta}\\
\frac{dY_{\deltaN}}{d x_{\SM}} 
	& = \frac{x_{\SM} s_{\SM}}{H(m_\etap)}~\langle \sigma v \rangle_{\etap\etap \to  \deltaN \deltaNb}~\Big[Y_{\etap}^2 - Y_{\deltaN}^2 Y_{\etap,\eq}^2/Y_{\deltaN, \eq}^2\Big]\,, \label{eqn:boltzdelta}
\end{align}
\end{subequations}
in which we have included for completeness the inverse processes in Eq.~\eqref{eqn:boltzdelta} neglected in Eq.~\eqref{eqn:expydelta}, 
and we have applied the usual detailed balance relations.

While in practice this coupled system must be solved simultaneously, it is instructive to first examine in detail the behavior of the pure $\etap$ evolution, assuming that $Y_{\deltaN}$ can be neglected. 
That is, the simplified system
\begin{equation}
	\frac{dY_{\etap}}{d x_{\SM}} \simeq - \frac{x_{\SM} s_{\SM}}{H(m_{\etap})}  \langle \sigma v \rangle_{2\etap \to 4\etap} \frac{Y^2_{\etap}}{Y^2_{\etap,\eq}}\Big[Y^2_{\etap} - Y^2_{\etap,\eq}\Big]\,.
\end{equation}
As in Sec.~\ref{sec:deltaY}, the equilibrium $\etap$ yield $Y_{\etap,\eq} = \rs n_{\etap,\eq}(x_d)/s_d(x_d)$, and can therefore be expressed purely as a function of $x_d$, 
as can the thermally averaged $2\etap \to 4\etap$ cross-section, explicitly presented above in Eq.~\eqref{eqn:sigmavetap}. 
In particular, as $x_d \to \infty$
\begin{equation}
	\label{eqn:Yetaeq}
	Y_{\etap,\eq} \simeq \frac{r_s}{x_d +1}\,,
\end{equation}
while $\langle \sigma v \rangle \sim e^{-2x_d}$. 
Hence the collision term of the Boltzmann equation nonetheless vanishes for $x_d \to \infty$. 
That is, $Y_{\etap} \to \text{const}$ for $x_d \to \infty$ is a solution, and the $\etap$ undergo a freeze-out.

The freeze-out condition $Y_{\etap} - Y_{\etap,\eq} \simeq Y_{\etap,\eq}$ implies that freeze-out occurs once
\begin{equation}
	\label{eqn:etapfocondition}
	\frac{dY_{\etap,\eq}}{d x_{\SM}}  \simeq -2 \frac{x_{\SM} s_{\SM}}{H(m_{\etap})}\langle \sigma v \rangle_{2\etap \to 4\etap}Y^2_{\etap,\eq}\,.
\end{equation}
Anticipating that the freeze-out typically occurs once the $\etap$ are non-relativistic,
we may directly express $x_{\SM}$ in Eq.~\eqref{eqn:etapfocondition} in terms of $x_d$ via the non-relativistic entropy ratio relation~\eqref{eqn:entropyratioNR} and the relation~\eqref{eqn:rsxiinf}.
Including the explicit form of the thermal cross-section~\eqref{eqn:sigmavetap}, the freeze-out then occurs at
\begin{equation}
	\label{eqn:etapxfo}
	x_{d,f} \simeq -\frac{1}{7}W_{-1}\bigg[-2 \times 10^3 \frac{\Lambda^3 N^{61/2} g_{*,\SM}^{3/2}}{\zeta^3 \xi^6 M_{\text{pl}}^3}\bigg]\,,
\end{equation}
where $W_{-1}$ is the product logarithm function on the negative real axis.
For typical values of $N \sim 10$, $\Lambda \sim 0.1$\,GeV and $\xiinf \sim 10^{-2}$, this corresponds to $x_{d,f} \sim 10$.
Since the product logarithm encodes only a weak dependence on $\Lambda$ and $N$,
with reference to Eq.~\eqref{eqn:Yetaeq} and the relation~\eqref{eqn:rsxiinf} we obtain a simple power law behavior for the relic abundance
\begin{equation}
	\label{eqn:ometa}
	\Omega_{\etap}h^2 \sim 0.1 \frac{s_0}{\rho_c} \Lambda N^{3/2} \xiinf^3
\end{equation}	
That is, we expect a power law behavior for the DM contour
\begin{equation}
	\label{eqn:etappowlaw}
	\Lambda \sim N^{-3/2}\,.
\end{equation}

Combined with the result that $\Omega_{\deltaN}h^2 \sim e^{-2(c_*+1)N}$ from Sec.~\eqref{sec:deltaY}, 
this approximate analysis allows us to develop intuition for the interplay between $\deltaN$ and $\etap$ abundances. 
As $\Lambda$ grows along the fixed-$N$ DM contour for pure $\deltaN$ dark matter, at some point $\Lambda$ becomes sufficiently large that the $\etap$ abundance becomes important.
The requirement $\Omega_{\deltaN} + \Omega_\etap \le \Omega_{\text{DM}}$ then pushes the DM contour to larger $N$ in order to suppress the $\deltaN$ contribution.
The exponential suppression quickly depletes the $\deltaN$ abundance, such that the DM contour then rapidly transitions to the power law in Eq.~\eqref{eqn:etappowlaw}.
In the next section, we show explicit numerical results that confirm this behavior.

\section{Numerical results}
\label{sec:numstudy}
\subsection{Numerical implementation}
In the previous section, we presented an approximate, quantitative analysis for computing the $\etap$ and $\deltaN$ relic abundances. 
While these analyses describe the qualitative behavior of the $\deltaN$ and $\etap$ coupled system,
for benchmark studies we instead use numerical methods to compute the $\etap$ and $\deltaN$ abundances, just as in Sec.~\ref{sec:numxsec}.
We first provide pertinent details of the implementation of these methods, then proceed to the full results for various benchmarks.

\subsubsection{Boltzmann equation}
The fully-coupled Boltzmann equations describing the evolution of the comoving number densities of the $\etap$ and $\deltaN$ are given in Eq.~\eqref{eqn:boltzsystem}.
For ease of controlling numerical errors, we convert the differential equation for $Y_{\etap}$ into an equation for $L_{\etap} = \ln(Y_{\etap})$. Thus, the equations we solve are
\begin{subequations}
\label{eqn:boltznum}
\begin{align}
	\dv{L_{\etap}}{\ln(x_{\SM})}
	 & = - \sqrt{\frac{\pi}{45}}M_{\text{pl}}\,g^{1/2}_{\text{eff},*} T_\SM s^2_{\SM} \expval{\sigma_{4\etap \to 2\etap} v^3} e^{L_{\etap}}\qty[e^{2L_{\etap}} - e^{2L_{\etap,\text{eq}}}] \,,\\
	\dv{Y_{\deltaN}}{x_{\SM}}
	& = \sqrt{\frac{\pi}{45}}\frac{M_{\text{pl}}\,g^{1/2}_{\text{eff},*}}{x_\SM}\expval{\sigma_{\etap\etap \to  \deltaN \deltaNb} v } e^{2L_{\etap}}\,,
\end{align}
\end{subequations}
where
\begin{align}
    g^{1/2}_{\text{eff},*} \equiv \qty(1 + \dfrac{T_{\mathrm{SM}}}{3h_\SM}\dv{h_\SM}{T_\SM})\dfrac{h_\SM}{\sqrt{g_\SM}}\,,
\end{align}
Because the SM-dominated evolution $dt/dT_{\mathrm{SM}}$ is a well-studied function of $T_{\mathrm{SM}}$, we solve these coupled equations with respect to $x_{\mathrm{SM}}$.
This is very different to the discussion of Sec.~\ref{sec:darkcos}, in which $x_d$ was the natural choice to characterize the dark sector dynamics.
We implement $g^{1/2}_{\mathrm{eff},*}$ using the results of Drees \emph{et.~al.}~\cite{drees2015effects}.

To solve these equations, we use the variable order, implicit, stiff-ODE solver \texttt{radau}~\cite{hairer1999stiff},\footnote{Available from \url{https://unige.ch/~hairer/software.html}.} 
converted from \texttt{FORTRAN} to \texttt{C++} using \texttt{f2c}~\cite{feldman1990fortran}.
We assume that the phase-transition occurs -- and our numerical evolution begins -- at a dark temperature $T_{d} \sim \Lambda / 2$, 
so that the $\Delta$ will in general be very cold just after the dark sector confining phase-transition. 
We thus assume always that the initial $\Delta$ abundance is zero. 
This allows use to neglect the back-reactions between the $\etap$ and $\deltaN$, i.e. the $\deltaN\bar{\deltaN}\to\etap\etap$ collision term.

The discussion in Sec.~\eqref{sec:darkcos} also relied on various assumptions, such as energy domination by the SM sector \eqref{eqn:SMdom} and imposing the non-relativistic limit for the $\etap$. 
We now relax and/or test these assumptions.

\subsubsection{Dark temperature}
The entropy ratio conservation relation~\eqref{eqn:entcons} can be written as
\begin{equation}
	\xi^3 \simeq \frac{h_{\SM}(T_\SM)}{h_{\SM,\infty}}\frac{h_{d,\infty}}{h_{d}(\xi T_\SM)}\xiinf^{3}\,,
\end{equation}
where the effective entropic number of degrees of freedom stored in the dark sector,
\begin{equation}
	h_d(T_d) \sim \sum_{i=\etap,\deltaN}\frac{45g_{i}}{4\pi^4}\,x_{d,i}^3\, \sum_{n=0}^{\infty}\dfrac{\eta_{i}^{n}}{(1+n)}K_{3}[(1+n)x_{d,i}]\,,\qquad \eta_{\etap,\deltaN} = \pm1.
\end{equation}
One may derive upper and lower bounds on $\xi$ from the asymptotic behavior of $h_{\mathrm{SM}}$ and $h_{d}$, yielding
\begin{equation}
    \bigg[\frac{h_\SM(T_\SM)h_{d,\infty}}{(7g_{\deltaN}/8+g_{\etap})h_{\SM,\infty}}\bigg]^{1/3}\xiinf< \xi(T_\SM) < \frac{2x_\SM}{W\big(2x_\SM/D^2\big)}\,,
\end{equation}
where $g_{\etap}$ and $g_{\deltaN}$ are the internal degrees of freedom of the $\etap$ and $\deltaN$,
\begin{equation}
    D = g_{\eta}\frac{h_{\SM}(T_\SM)h_{d,\infty}}{h_{\SM,\infty}}\bigg(\frac{45x_\SM^{5/2}}{4\sqrt{2}\pi^{7/2}}\bigg)^{-1}\xiinf^3\,,
\end{equation}
and $W(x)$ is again the Lambert function. 
Given these initial bounds, which bracket the true value of $\xi$, one may then use a simple bisection algorithm to numerically determine $\xi$ for a given SM temperature and initial $\xiinf$.

\subsection{Benchmark results}
In order to examine the numerical solutions of the Boltzmann system~\eqref{eqn:boltznum} in the $N$--$\Lambda$ plane, we must choose benchmark values 
for the momentum expansion parameters of the $\etap$ Lagrangian, $\lambda_{1,2}$, as well as the early temperature ratio $\xiinf$ and the effective exponent $c_*$. 
Since we expect that $\lambda_k \sim 1$, a `minimal' benchmark is simply to take $\lambda_1 = 1$ and $\lambda_2 = 0$. 
This fully correlates the $\etap$ freeze-out $2\to4$ and self-interaction $2\to2$ cross-sections.
Based on expectations from $\deltaN_{\text{eff}}$ constraints in Sec.~\ref{sec:cons}, we choose an initial benchmark value for the temperature ratio $\xiinf = 10^{-2}$.
For this and all other benchmarks below we fix the value of $c_* \sim 1$ from the results of Fig.~\ref{fig:deltaDM}, such that the $\deltaN$ DM contour sits on $N = 7$ as is thus physically permitted.
A summary of this and other benchmarks is shown in Table~\ref{tab:benchmarks}.

\newcolumntype{Y}{ >{\centering\arraybackslash $} m{2cm} <{$} }
\begin{table}[t]
	\renewcommand{\arraystretch}{1.2}
	\centering
	\begin{tabular}{@{\extracolsep{\fill}}cYYY}
		\hline\hline
		Benchmark & \lambda_1  & \lambda_2 & \xiinf\\
		\hline
		Minimal & 1 & 0 & 10^{-2} \\
		Decorrelated & 0.1 & 1 & 10^{-2}\\
		Hot & 0.1 & 1 & 5 \times 10^{-2} \\
		Low SI & 10^{-3} & 1 & 10^{-2} \\
		\hline\hline
	\end{tabular}
	\caption{Benchmarks for the momentum expansion parameters of the $\etap$ Lagrangian, $\lambda_{1,2}$, and the early temperature ratio $\xiinf$. }
	\label{tab:benchmarks}
\end{table}

\begin{figure}[t]
    \centering
    \begin{subfigure}[b]{0.5\textwidth}
        \centering
        \includegraphics[width=\textwidth]{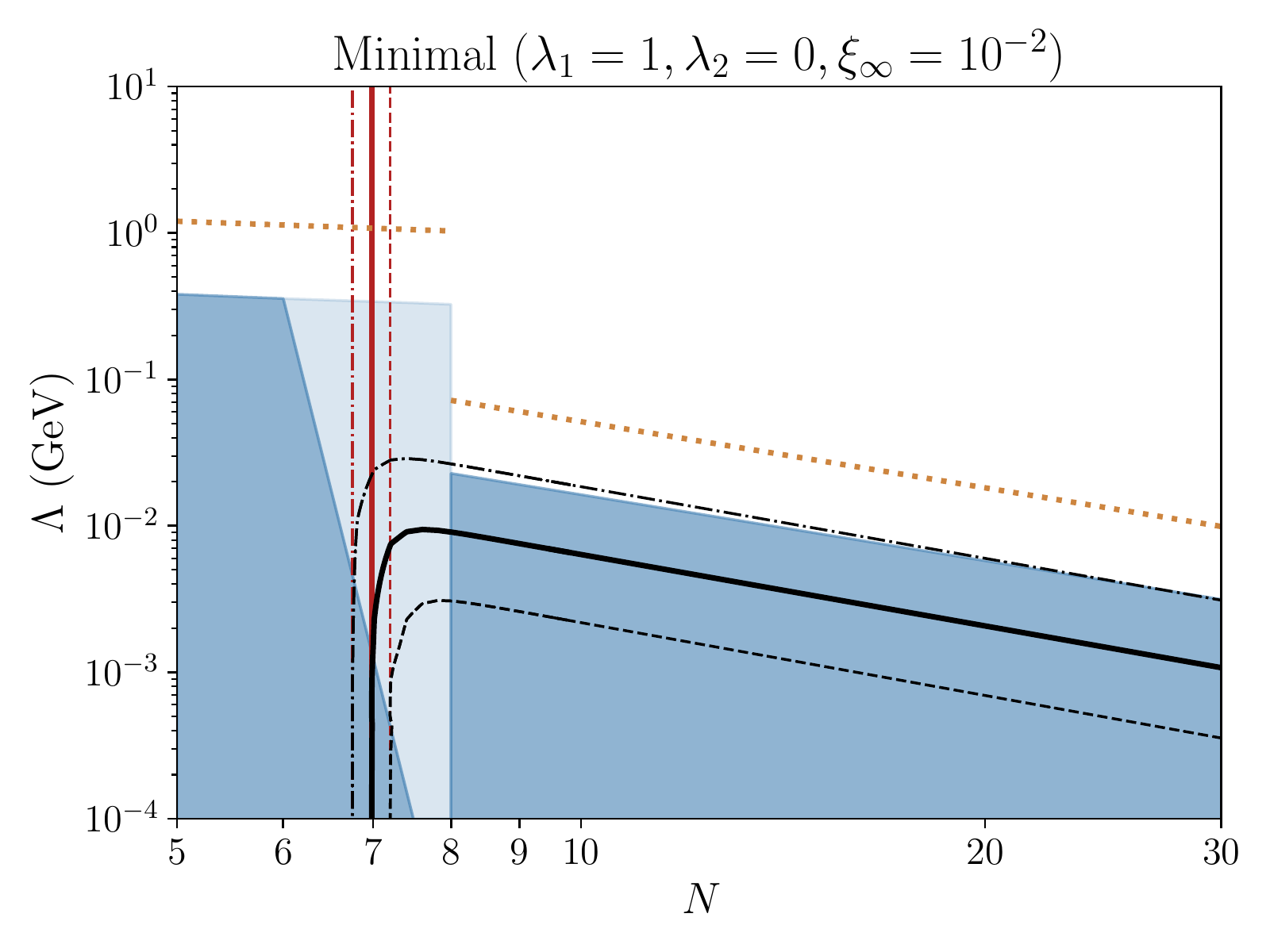}
        \caption{}\label{fig:LNminimal}
    \end{subfigure}\hfill
    \begin{subfigure}[b]{0.5\textwidth}
        \centering
        \includegraphics[width=\textwidth]{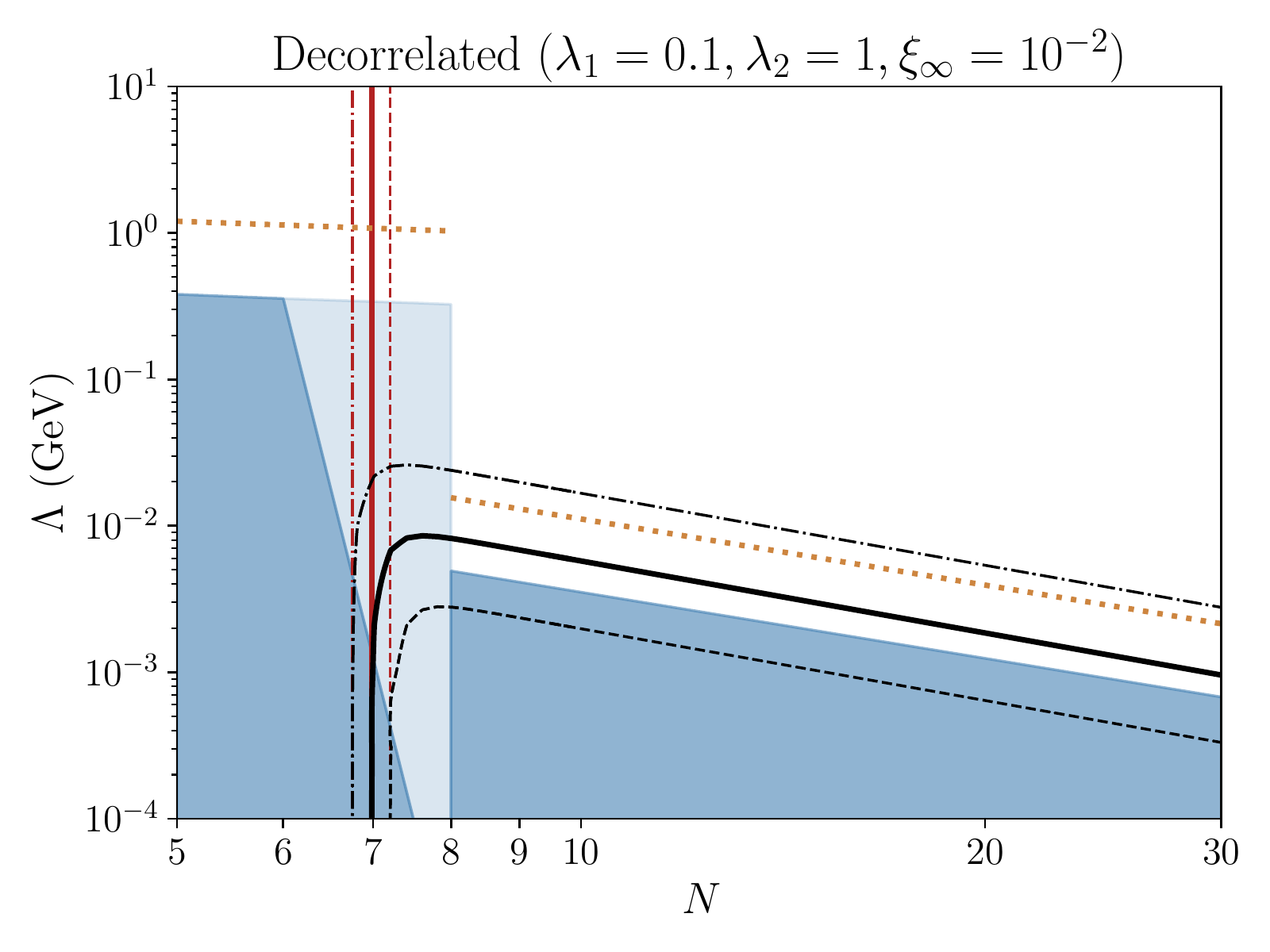}
        \caption{}\label{fig:LNdecorrelated}
    \end{subfigure}
    \begin{subfigure}[b]{0.5\textwidth}
        \centering
        \includegraphics[width=\textwidth]{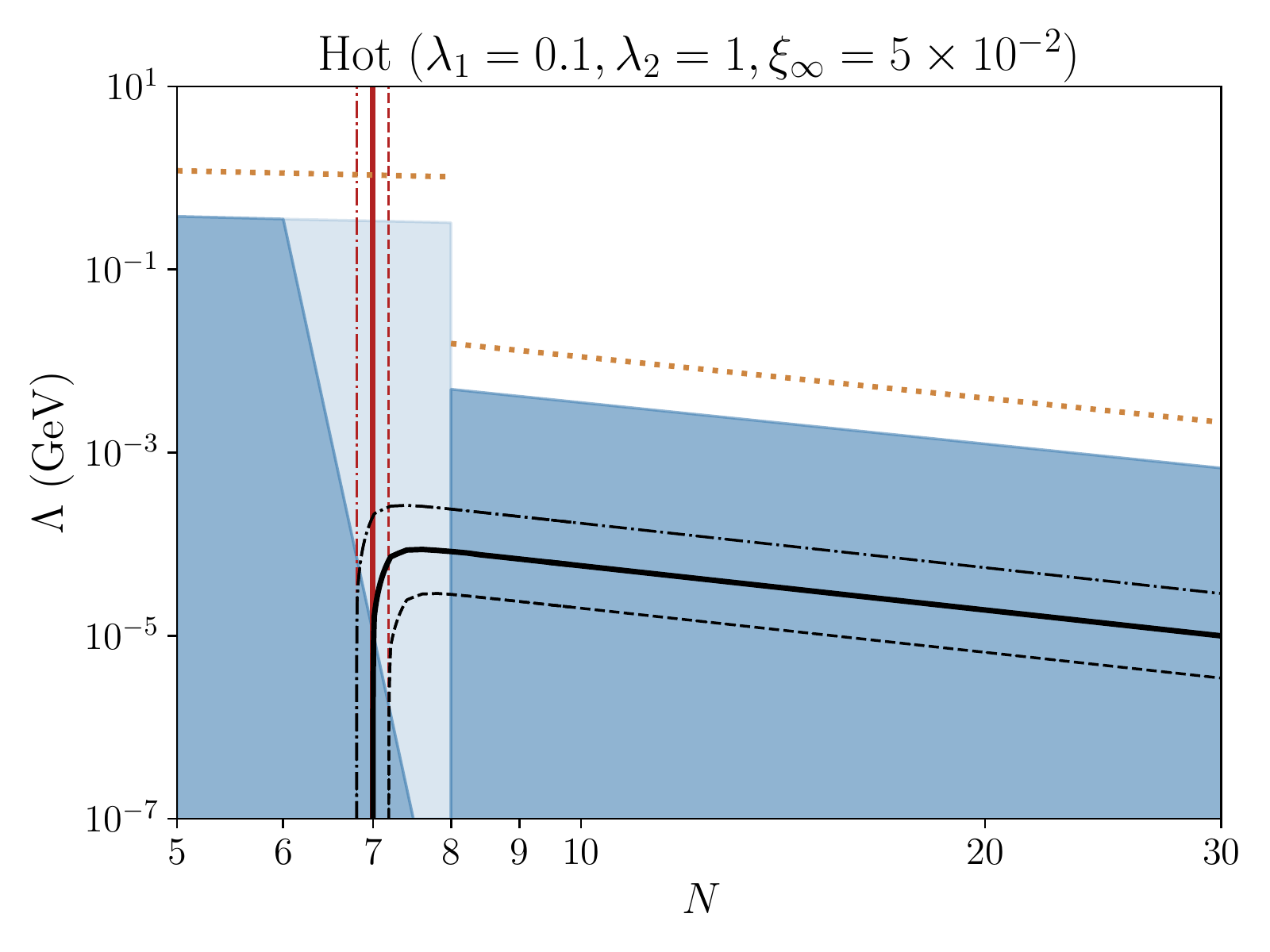}
        \caption{}\label{fig:LNhot}
    \end{subfigure}\hfill
    \begin{subfigure}[b]{0.5\textwidth}
        \centering
        \includegraphics[width=\textwidth]{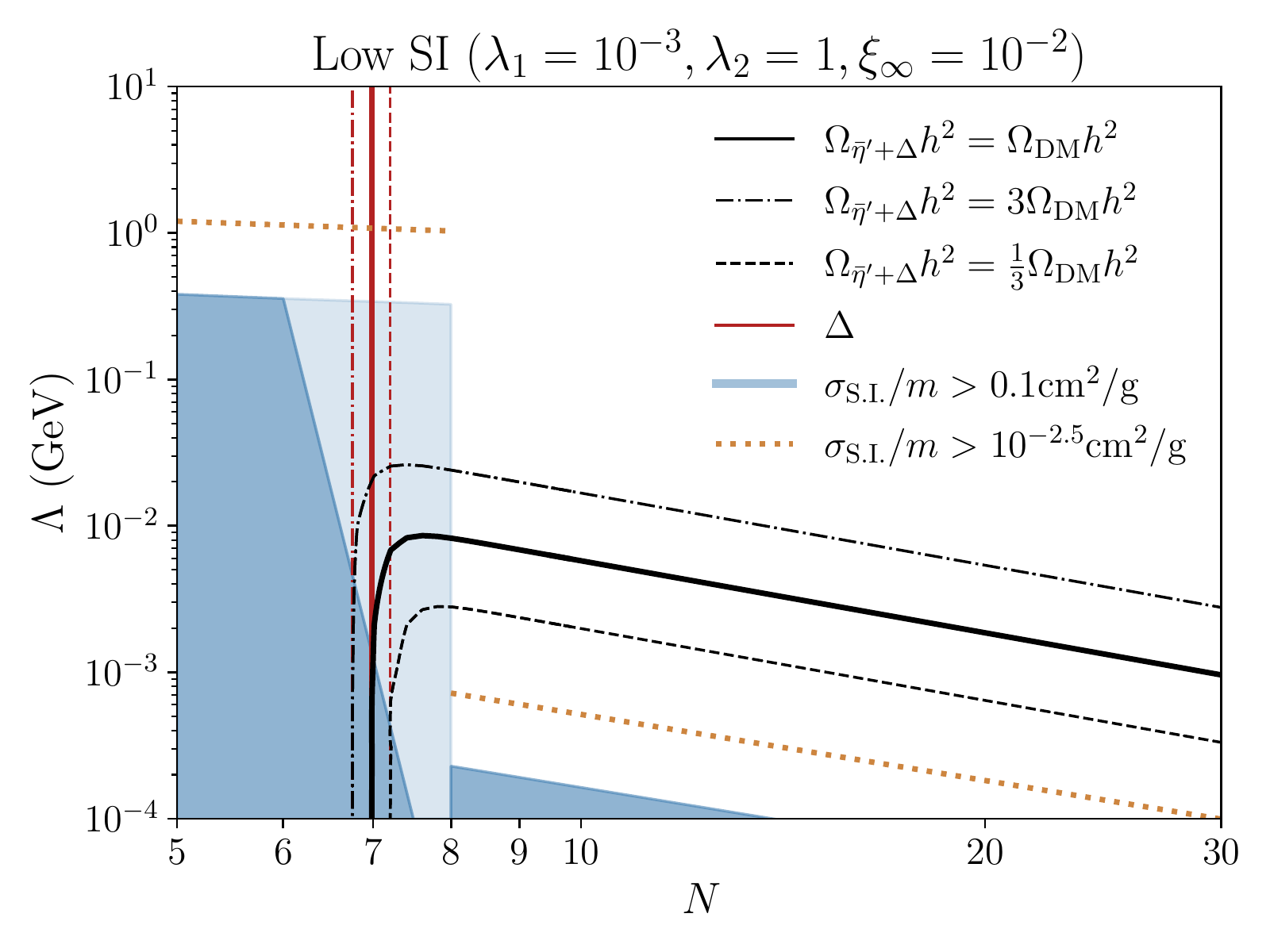}
        \caption{}\label{fig:LNlowSI}
    \end{subfigure}
    \caption{The $\deltaN$-$\etap$ DM contour (black) in the $N$-$\Lambda$ plane, that corresponds to $\Omega_{\deltaN}h^2 + \Omega_{\etap}h^2 = \Omega_{\text{DM}}h^2$.
    Also shown contours for an overabundance (dot-dashed) and underabundance (dashed) of DM. 
    Single species (extrapolated) self-interaction bounds are shown in solid (light) blue, for the $\deltaN$ and $\etap$ on the left and right, respectively. See text for details.
    Dashed orange lines show the extent of future self-interaction bounds $\sigma_{\text{SI}}/m < 10^{-2.5}$\,cm$^2/$g.
    The pure $\deltaN$ DM contours are shown in red. Note the different scale on `Hot' benchmark $\Lambda$ axis.
    }\label{fig:LN}
\end{figure}

In Fig.~\ref{fig:LNminimal} we show the contour in the $N$-$\Lambda$ plane that corresponds to the two-component system $\Omega_{\deltaN}h^2 + \Omega_{\etap}h^2 = \Omega_{\text{DM}}h^2$.
As expected from the discussion in Sec.~\ref{sec:etadeltaDM}, for low $N$ the DM is dominated by $\deltaN$ resulting in a $\Lambda$-independent vertical contour. 
As $\Lambda$ increases along this contour, the $\etap$ $2 \to 4$ interaction eventually becomes sufficiently weak and the $\etap$ mass becomes sufficiently large, 
such that a significant $\etap$ relic to be produced. 
This causes a turn over to a power-law like behavior for large $N$, with the DM dominated by the $\etap$. 
The power law is numerically $\Lambda \sim N^{-1.6 \pm 0.05}$, close to our expectation from the analysis in Sec.~\ref{sec:etadeltaDM}.
The turn-over itself contains non-negligible $\deltaN$ and $\etap$ populations: a `mixed regime'.
To guide intuition, we also show contours for an over and underabundance of DM. 
Though $N$ must be an integer, we plot it as a continuous variable, since variation in $c_*$ or other nuisance parameters can move the vertical portion of contour.
(The physical value of $c_*$ or other nuisance parameters may imply that pure $\deltaN$ DM falls on a non-integral value and is thus unphysical, 
while instead a single point on the mixed regime turn-over at $N=7$ is physical.)
The relic abundance contours for pure $\deltaN$ DM, in the case that the $\etap$ is e.g. unstable, are shown in red (see Sec.~\ref{sec:deltaDM}).
For $\Lambda \gtrsim 0.3$\,GeV, the self-interaction constraints on pure $\deltaN$ DM are relaxed.

In the regime for which the DM abundance is either pure $\deltaN$ or pure $\etap$, we show the corresponding pure $\deltaN$ and pure $\etap$ self-interaction bounds by solid blue regions. 
The relevant self-interaction bounds in the mixed regime, with non-negligible $\deltaN$ and $\etap$ populations, do not interpolate simply between these two regimes. 
As a conservative estimate, we instead extrapolate the pure $\deltaN$ and pure $\etap$ self-interaction bounds into this regime, showing this extrapolation by light shading. 
This transition regime is very sharp in $N$, but may extend over an order of magnitude in $\Lambda$.

One sees in Fig.~\ref{fig:LNminimal} that this minimal benchmark is excluded by the self-interaction bounds, 
no matter where one is on the two-component DM contour.
Since the $2 \to 4$ cross-section scales as $|10\lambda_1^2 + \lambda_2|^2$, 
a `decorrelated' benchmark choice $\lambda_1 \lesssim 0.1$ and $\lambda_2 \sim 1$ instead decorrelates the parametrics of the freeze-out and self-interaction processes. 
We thus choose a second benchmark $\lambda_1 = 0.1$ and $\lambda_2 = 1$, shown in Fig.~\ref{fig:LNdecorrelated}. 
While the pure $\deltaN$ DM regime of the contour is still excluded, this benchmark falls in the allowed region for pure $\etap$ DM. 
This benchmark may, however, be probed by future self-interaction constraints at the $\sigma_{\text{SI}}/m <10^{-2.5}$cm$^2$/g level, shown by dashed orange lines.

In order to characterize the sensitivity to $\xiinf$ (and anticipating a discussion of the effects of the $\delta N_{\text{eff}}$ bounds and the range of validity of the assumption of SM energy domination)
we also consider a `hot' version of the decorrelated benchmark, with $\xiinf = 5\times 10^{-2}$.
The hotter dark sector implies more entropy in the dark sector, requiring a lighter $\etap$ in order not to overgenerate the DM. 
This pushes the $\etap$ DM contour to lower $\Lambda$, resulting in more severe exclusion by self-interaction bounds.

Finally, we consider a benchmark for which the self-interaction is turned-off -- a `low SI' benchmark -- by taking $\lambda_1 \lll 1$ while $\lambda_2 =1$.
In this scenario, the entire pure $\etap$ DM regime of the contour is allowed; the pure $\deltaN$ regime is excluded, but a mixed population of $\deltaN$ and $\etap$ might be allowed. 
The latter requires a detailed study of self-interaction bounds for two-component DM, beyond the scope of this work.

\subsection{$\delta N_{\text{eff}}$ constraints}
\begin{figure}[t]
        \centering
        \includegraphics[width=7.5 cm]{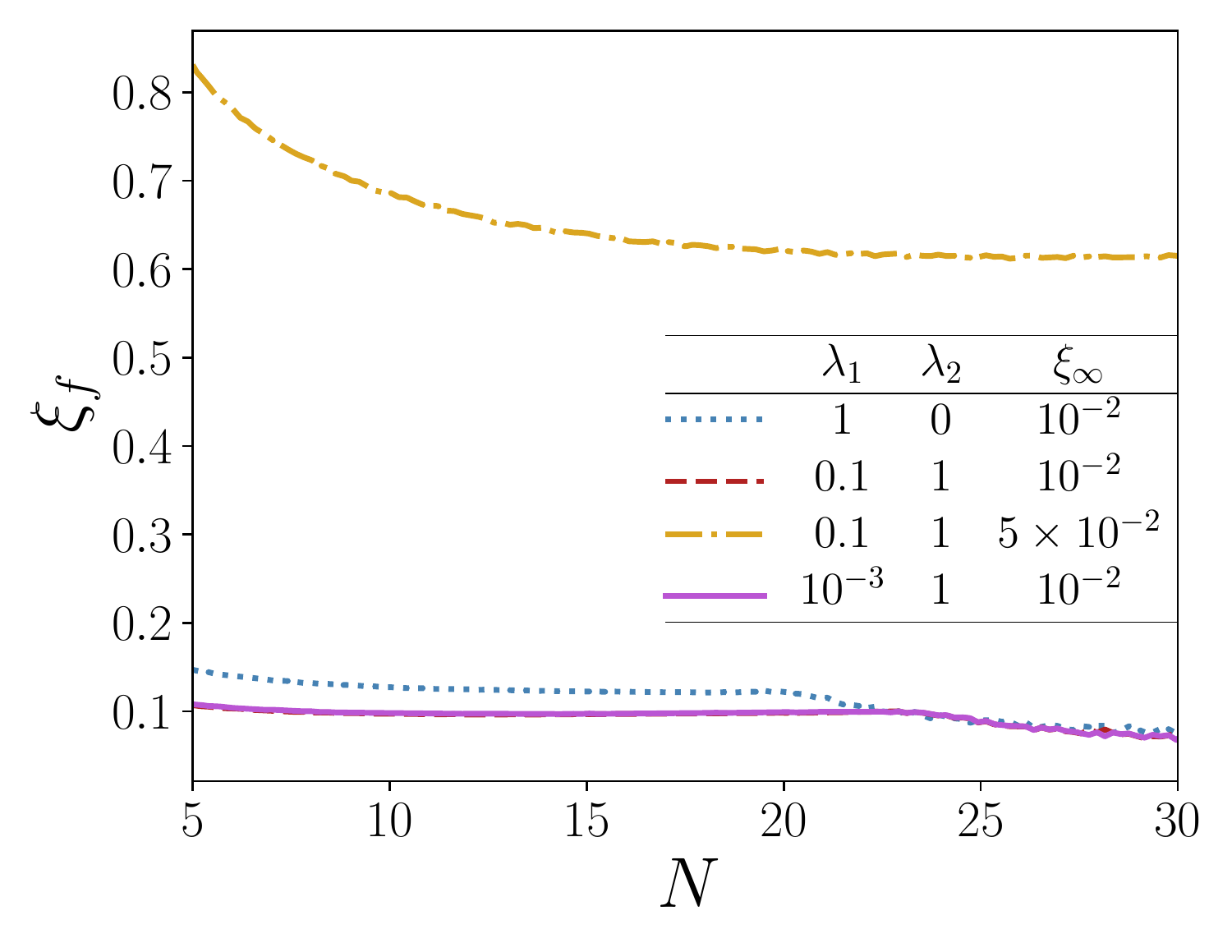}\hfill
          \includegraphics[width=7.5 cm]{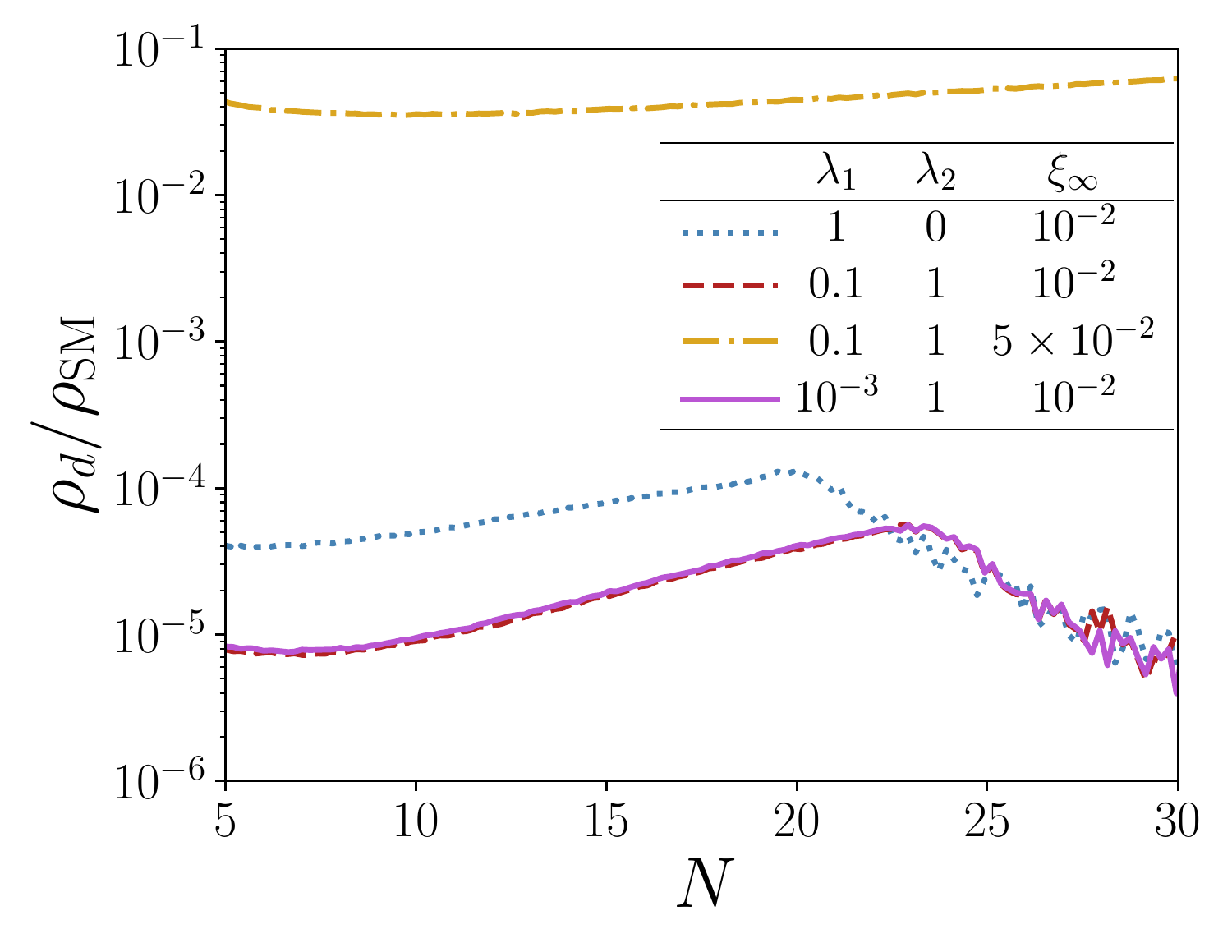}
        \caption{Left: The ratio of the dark-SM temperature, $\xi_f$, at $\etap$ freeze-out. Right: The dark-SM energy density ratio at freeze-out.}
        \label{fig:xifo}
\end{figure}

\begin{figure}[t]
        \centering
        \includegraphics[width=0.45\textwidth]{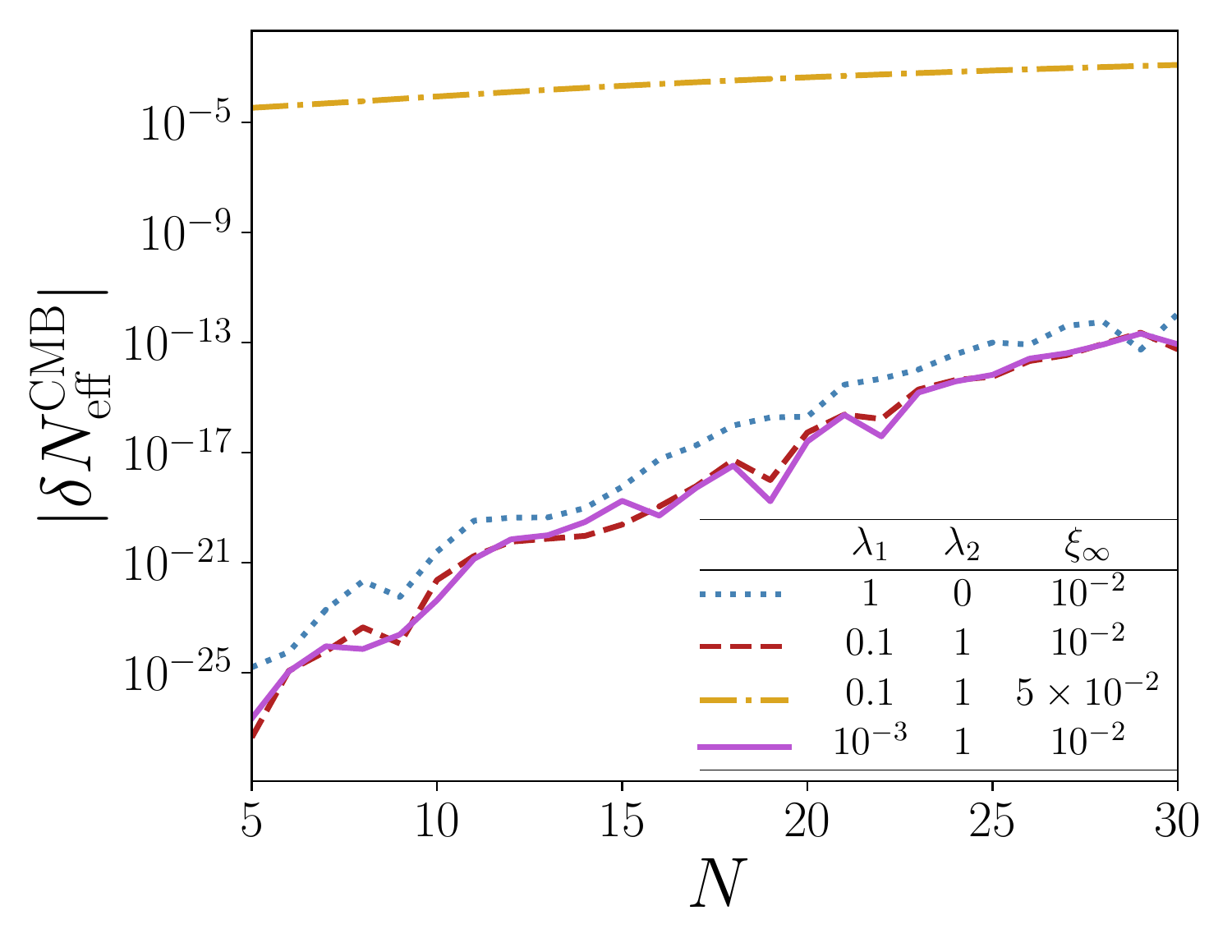}
        \includegraphics[width=0.45\textwidth]{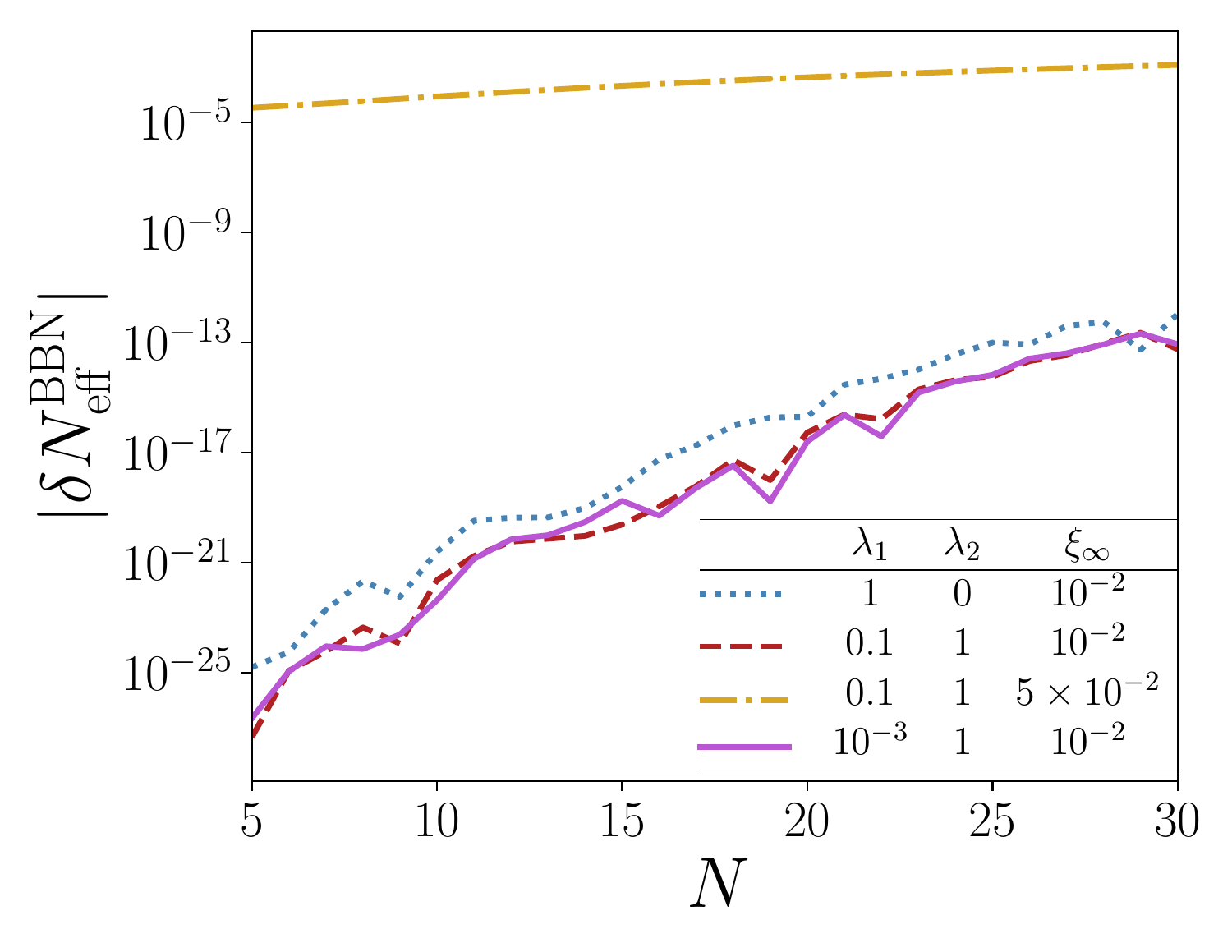}
        \includegraphics[width=0.45\textwidth]{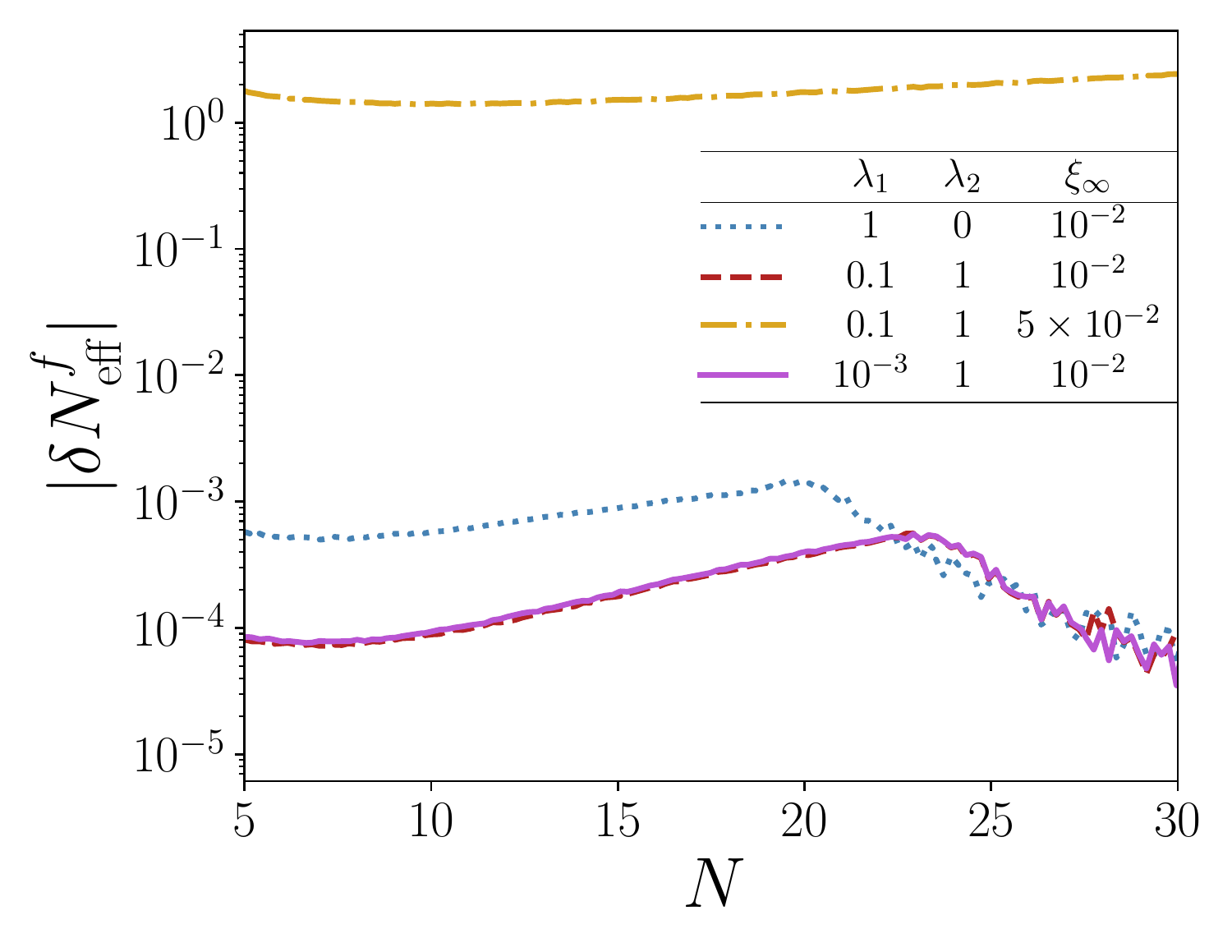}
        \caption{The shift in effective degrees of freedom $\delta N_{\mathrm{eff}}$; at CMB (top left), BBN (top right) and at $\etap$ freeze-out (bottom).}
        \label{fig:neff}
\end{figure}

For all these benchmarks, it is important to also check if they satisfy the relevant $\delta N_{\text{eff}}$ bounds discussed in Sec.~\ref{sec:cons}.
Throughout the cosmological evolution of the dark sector, the highest dark-SM temperature ratio typically occurs at $\etap$ freeze-out, i.e., for any benchmark $\xi \le \xi_f$, 
the temperature ratio at freeze-out.
In Fig.~\ref{fig:xifo}, for each benchmark we show the freeze-out temperature and energy density ratios on the two-component DM contour 
-- the solid black contours shown in Figs.~\ref{fig:LN} -- as a function of $N$.
For the three benchmarks with $\xiinf = 10^{-2}$, the freeze-out ratio is $\mathcal{O}(10^{-1})$, so that the energy density ratio at freeze-out $\lesssim 10^{-4}$.  
The corresponding $\delta N_{\text{eff}}$ at either BBN or CMB is therefore negligible, as shown in Fig.~\ref{fig:neff}.
(At $N\sim20$ the $\etap$ freeze-out happens to occur contemporaneously with the SM QCD phase transition, 
generating a moderate increase in the SM sector temperature and thus a dip in $\xi_f$.)

For the hot benchmark, with $\xiinf = 5 \times 10^{-2}$, in Fig.~\ref{fig:xifo} we see a marked increase to $\xi_f \sim 1$, and a correspondingly larger freeze-out energy density ratio $\sim 10^{-1}$.
This corresponds to $\delta N _{\text{eff}} \sim 1$ at freeze-out, as shown in Fig.~\ref{fig:neff}, 
but subsequent redshifting~\eqref{eqn:Tdredshift} results in $\delta N _{\text{eff}} \lesssim 10^{-2}$ at either BBN or CMB, safely within current constraints.
Similar to the discussion in Sec.~\ref{sec:cons}, we thus see that one requires typically $\xiinf \lesssim \text{few}\times10^{-2}$ to safely satisfy the $\delta N _{\text{eff}}$ bounds.
Moreover, up to $\xiinf \sim \text{few }\times 10^{-2}$, the SM dominates the energy budget over the whole cosmological evolution, as expected from the discussion in Sec.~\ref{sec:thermal}.
As shown above, the hot benchmark is also heavily excluded by self-interaction constraints: 
Increasing $\xiinf$ thus intrudes on both the self-interaction and $\delta N_{\text{eff}}$ constraints, 
though one expects which constraint is tighter in practice to depend upon the details of the particular benchmark. 

\section{Summary and outlook}
\label{sec:summ}
We entertained and studied a scenario in which the dark matter belongs to a completely secluded dark sector, 
featuring a single flavor of light vector-like dark quarks charged under a confining $SU(N)$ gauge group.
Theoretical results for the large-$N$ limit imply that below the confinement scale, $\Lambda$, the sector features two stable bound states:
A light quark-antiquark $\etap$ state, analog to the $\eta^\prime$ SM meson, with mass $\sim \Lambda/\sqrt{N}$; 
and a heavy baryonic $\deltaN$ state, the analog of the $\Delta^{++}$ SM baryon, with mass $\sim \Lambda N$.
Absent a portal between the Standard Model and this dark sector, the two sectors give rise to two uncorrelated thermal baths, with two distinct temperatures,
whose thermal history is characterized by the asymptotic, large-temperature ratio of the dark-SM sector temperatures, $\xiinf$.

The cosmological abundance of the two stable species is controlled by two qualitatively different mechanisms: 

On the one hand, large-$N$ arguments imply that the interactions of the Boltzmann-suppressed $\deltaN$ with the $\etap$ plasma are exponentially suppressed in $N$, 
such that $\deltaN$'s are produced out of thermal equilibrium  -- i.e. freeze-in --  from the $\etap$ thermal bath via $\etap\etap\to\deltaN\bar\deltaN$. 
The resulting cosmological abundance $\Omega_\deltaN\sim e^{-2(c_* +1)N}$, with $c_*$ an ${\cal O}(1)$ parameter. 
Importantly, this is independent of $\Lambda$ and produces a DM relic abundance for $N\lesssim10$. 

On the other hand, the $\etap$ abundance is determined by $4 \leftrightarrow 2$ cannibalization and freeze-out.
The self-interactions of the $\etap$ are described by higher-order derivative interactions of the trivial $\etap$ chiral Lagrangian. 
Applying large-$N$ scaling arguments, 
the $\etap$ dynamics is predominantly characterized by just the two parameters, $\lambda_{1,2}$, entering the amplitudes of $2\leftrightarrow2$ and $2\leftrightarrow4$ $\etap$ processes, respectively.
This results in a highly-predictive relationship between the $2 \leftrightarrow 2$ self-interactions and the $4 \leftrightarrow 2$ cannibalization and freeze-out.
Approximate expressions for the $2\leftrightarrow2$ and $2\leftrightarrow4$ cross-section as a function of the dark sector temperature, $N$, $\Lambda$, and $\lambda_{1,2}$
allowed us to develop a detailed, approximate analysis of the cosmological evolution of the $\etap$, with the simple result that the relic density of $\etap$ scales as $\Lambda \sim N^{-3/2}$.
Taken together, the two different production mechanisms mean that each of the two species constrains the allowed range of $\Lambda$ and $N$ for the two-component $\etap$-$\deltaN$ DM system,
resulting in a $\etap$-$\deltaN$ DM contour with regimes of either mostly $\deltaN$, mostly $\etap$ DM, or a small mixed regime where both populations are present.
This behavior is confirmed by our numerical studies.

The scenario under consideration is further constrained by the effects of DM self-interactions on dark matter halos. 
In turn, these are controlled by the $\etap\etap\to\etap\etap$ cross section for the case of dominant $\etap$ DM, 
and by $\deltaN\deltaN\to\deltaN\deltaN$ scattering for the case of dominant $\deltaN$ DM: The intermediate case is more complicated to observationally constrain. 
We find that generally self-interaction constraints are extremely strong for the pure $\deltaN$ DM regime on the two-component $\etap$-$\deltaN$ contour.
The pure $\etap$ DM regime is also constrained by self-interactions, although in ways that depend on other parameters, in particular $\xiinf$ and $\lambda_1$.
Constraints also arise from $\etap$ contributions to the effective number of light relativistic degrees of freedom, $N_{\text{eff}}$, at the BBN and CMB epochs. 
We showed that these bounds typically imply that $\xiinf\lesssim \text{few}\times 10^{-2}$. This is a very small ratio, and nominally implies non-trivial physics in the ultraviolet.

For four benchmark choices of $\lambda_{1,2}$ and $\xiinf$, using numerical computations of the thermally-averaged cross sections, 
we studied the DM abundance contours and allowed regions in the $N$--$\Lambda$ plane (see Fig.~\ref{fig:LN}).
The most minimal scenario is excluded by self-interaction bounds.
While not guaranteed, improvements on constraints on dark matter self-interaction might very well produce evidence, or constrain, the other benchmarks under consideration.

We also considered briefly a scenario in which the accidental vector $U(1)_V$ symmetry is gauged, giving rise to $\etap$ decay to a dark photon. 
In this case, as long as the $\etap$ are long-lived enough, freeze-in of the $\deltaN$ is still possible, 
but the $\etap$ thermal bath does not have the opportunity to significantly exponentially heat with respect to the SM sector bath.
Moreover $\deltaN$ solely forms the DM relic. 
Since $\Omega_\deltaN h^2$ is independent of $\Lambda$, $\Lambda$ can be chosen over a very large range, 
such that $\deltaN$ may become very heavy and self-interaction constraints on the $\deltaN$ are alleviated,
while the $\etap$ is still long-lived enough and the dark photon coupling remains perturbative.
Finally, including a kinetic mixing portal with SM hypercharge would allow one to awake, as it were, from the nightmare scenario, 
by introducing the possibility of direct dark matter detection and of detection of new states with particle colliders.

\acknowledgments

DJR thanks Jamison Galloway, Duccio Pappadopulo, and Josh Ruderman for prior discussions of large-$N$ dark QCD, in the adjacent context of Forbidden Dark Matter~\cite{DAgnolo:2015ujb}.
We also thank Marat Freytsis, Simon Knapen, and Aneesh Manohar for helpful discussions and consultations.
LM and SP are partly supported by the U.S.\ Department of Energy grant number DE-SC0010107.
DJR is supported in part by the Office of High Energy Physics of the U.S. Department of Energy under contract DE-AC02-05CH11231.

\bibliographystyle{JHEP}
\bibliography{references}

\end{document}